\documentclass[aps,prb,twocolumn,groupedaddress]{revtex4}

\usepackage{graphicx}
\usepackage{dcolumn}
\usepackage{bm}
\usepackage{booktabs}
\usepackage{amssymb}
\usepackage{amsmath}
\usepackage{amsfonts}
\usepackage{scrextend}
\usepackage{enumerate}
\usepackage{hyperref}

\begin{document}

\title{On-the-fly machine learning force field generation:\\
       Application to melting points}

\author{Ryosuke Jinnouchi}
\affiliation{%
 University of Vienna, Department of Physics, Sensengasse 8/16, Vienna, Austria\\
}%
\affiliation{%
Toyota Central Research and Development Laboratories, Inc., 41-1 Yokomichi, Nagakute, Aichi 480-1192, Japan\\
}%
\author{Ferenc Karsai}
\affiliation{
VASP Software GmbH, Sensengasse 8, Vienna, Austria\\
}%
\author{Georg Kresse}
\affiliation{%
 University of Vienna, Department of Physics, Sensengasse 8, Vienna, Austria\\
}%

\begin{abstract}
An efficient and robust on-the-fly machine learning force field method is developed and integrated into an electronic-structure code. This method realizes automatic generation of machine learning force fields on the basis of Bayesian inference during molecular dynamics simulations, where the first principles calculations are only executed, when new configurations out of already sampled datasets appear. The developed method is applied to the calculation of melting points of Al, Si, Ge, Sn and MgO. The applications indicate that more than 99 \% of the first principles calculations are bypassed during the force field generation. This allows the machine to quickly construct first principles datasets over wide phase spaces. Furthermore, with the help of the generated machine learning force fields, simulations are accelerated by a factor of thousand compared with first principles calculations. Accuracies of the melting points calculated by the force fields are examined by thermodynamic perturbation theory, and the examination indicates that the machine learning force fields can quantitatively reproduce the first principles melting points.
\end{abstract}


\maketitle
\section{Introduction}
\label{section_introduction}
The quantitative prediction of first-order phase transitions of real materials from first principles (FP) calculations is a long-standing issue in condensed matter physics. The phase transition point is simply located at a temperature-pressure point where Gibbs free energies of two phases become identical. Its prediction is, however, quite challenging. Direct simulations using molecular dynamics (MD) and Monte Carlo methods are not computationally tractable because of the long time-scale nature of the phase transitions. Several approaches were proposed in order to solve this time-scale issue. One is an indirect approach on the basis of the thermodynamic integration method~\cite{1,2,3}. In this approach, the Gibbs free energy of a single phase is calculated by integrating a chemical potential derivative along a reversible thermodynamic path  that connects from a simple statistical model to a realistic interacting model. Another one is a direct approach, where the co-existence point of the solid-liquid interface is directly explored by MD~\cite{48,49,50,51,52,53}. There is an alternative approach~\cite{4} that transforms this out-of-equilibrium direct simulation to an equilibrium simulation by introducing a bias potential pinning the system in an interfacial state. In this third approach, the co-existence point is determined by using the free energy difference obtained from the mean force of the bias potential. However, all these methods need significant computational resources.

Recently developed machine learning force field (MLFF) techniques~\cite{5,6,7,8,9} have the potential to solve this problem. In those techniques, the potential energy of the system is described as a function of structural descriptors that map the 3$N$-dimensional structural information onto a lower-dimensional descriptor space, and the function is optimized to reproduce the FP data. High flexibility of the descriptors and this function allow for an accurate reproduction of the FP data, and the simulations are orders of magnitude faster using the generated force fields than the FP simulations.

However, applications of machine learning (ML) approaches are still limited to few simple materials. The difficulties are mainly in the force field generation process. Carefully selected reference datasets and a  tremendous amount of FP calculations on typically 2000-12000 structures~\cite{5,9,10,11,12,13} are needed in order to train force fields.
Furthermore, the data selection and parameter optimization are complex and involve a quite large number of trial and error steps.

On-the-fly force field generation~\cite{14,15,Jacobsen:prl2018,25} has the potential to overcome these limitations. In this method, energy, forces, and the stress tensor as well as their uncertainties are computed by Bayesian inference during an MD simulation. If the uncertainties are judged to be small, the computed energy, forces, and the stress tensor are used to integrate the equations of motions. If the uncertainties are judged to be large, FP calculations are executed in order to obtain new data that are then used to refine the force field. This error estimation and judgment step can realize efficient explorations of wide phase spaces and systematic data selection.

In this article, we present an efficient and robust algorithm that can be applied to liquid-solid phase transitions of a wide variety of materials. In Section~\ref{section_on_the_fly}, theories and equations used in our on-the-fly algorithm are presented. In Section~\ref{section_method_Tm}, method and parameters used in the simulations of the phase transitions are described. In section~\ref{section_Tm}, the calculated melting points are presented before we finally conclude the paper in section~\ref{section_conclusion}.
\section{Method: On-the-fly force field generation}
\label{section_on_the_fly}
In this section, after describing an outline of our on-the-fly MLFF generation scheme, the necessary methodologies composing this scheme are presented. For a concise presentation of the methodologies, we define \textit{structure datasets} and \textit{local configurations}. A single \textit{structure dataset} consists of the Bravais lattice, the atomic positions, the total energy, the forces and the stress tensor for one specific structure calculated by the FP method. We will label these datasets using the superscript $\alpha$.
For each atom in the structure, a \textit{local configuration} around this atom can be determined. This local configuration is mapped onto a set of descriptors describing the local environment around each atom as will be explained later on. Local structures and the central atom in a local structure are labeled using indices $i$ or $i_\mathrm{B}$.
Several structure datasets and local configurations are selected, and the ML force field is fitted to those. The selected datasets and configurations are referred to  as \textit{reference structure datasets} and \textit{local reference configurations}, respectively.
\begin{figure}
\includegraphics[width=0.40\textwidth,angle=0]{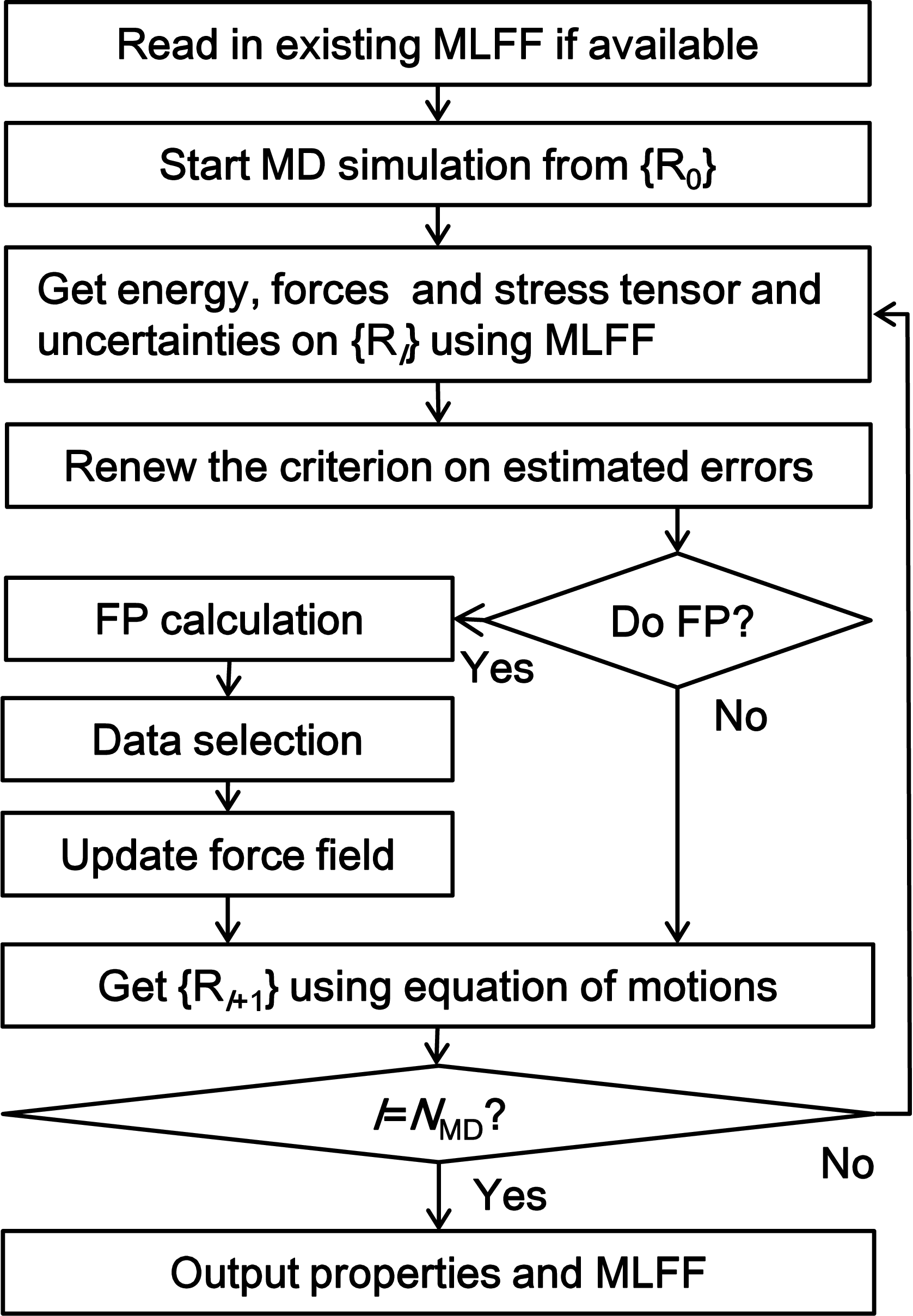}
\caption{Flowchart of our on-the-fly machine learning force field generation scheme.}
\label{fig1}
\end{figure}
\subsection{Outline of the on-the-fly force field generation}
\label{subsection_outline_of_on_the_fly}
Figure~\ref{fig1} shows the flowchart of our on-the-fly force field generation scheme. In our scheme, a force field is generated during MD simulations, as outlined below:
\begin{enumerate}[1)]
\item{ The machine predicts the energy, forces, stress tensor and their uncertainties on a given structure using the yet available force field.}
\item{ The machine decides whether to execute the FP calculation or not. The decision is done on the basis of the uncertainty in the prediction and a history of previous samplings. If the machine decides not to execute the FP calculations, the algorithm skips to step 5). Otherwise, it continues with step 3).}
\item{ The FP calculation is executed on the given structure, and the obtained structure dataset is stored as a candidate for a new reference structure dataset.}
\item{ If the number of the newly collected structures reaches a certain threshold, or if the uncertainty in the prediction becomes too large, the machine updates the set of reference structure datasets and local reference configurations and generates a new force field.}
\item{ Atomic positions and velocities are updated. If the machine judges that the force field is unreliable, the FP energy, forces and stress tensor are used. Otherwise those provided by the force field are used. Afterwards the machine returns to step 1) until the end of the MD simulation is reached ($I=N_{\mathrm{MD}}$ in Fig.~\ref{fig1}).}
\end{enumerate}
This scheme needs several key methodologies: an accurate description of the potential energy surface, an optimization of the parameters in the force field, evaluation of the uncertainty, setting of the threshold for the uncertainty,
 and sparsification and data selections. All these ingredients were implemented within the Vienna Ab initio Simulation Package (VASP)~\cite{16,17}. Their details are explained in the following subsections.
\subsection{Descriptor}
Our description of the potential energy surface is similar to that adopted in the Gaussian Approximation Potential (GAP)~\cite{6} with the Smooth Overlap of Atomic Positions (SOAP)~\cite{22} as a similarity measure. Several new features are, however, introduced to make the on-the-fly force field generation process more efficient and robust. In order to explain them in a concise manner, we formulate the energy and descriptor in this subsection. Here, we show for simplicity the equations only for single element systems, however, the extension to multi-element systems is straightforward.

\begin{figure}
\includegraphics[width=0.30\textwidth,angle=0]{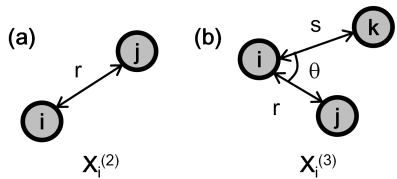}
\caption{(a) Radial and (b) angular descriptors.}
\label{fig2}
\end{figure}

In the method presented in this work, the potential energy $U$ of a structure with $N_{\mathrm{a}}$ atoms is approximated as a summation of local energies $U_{i}$ as
\begin{align}
U &= \sum\limits_{i=1}^{N_{\mathrm{a}}} U_i. \label{equation_pes}
\end{align}
Each local energy $U_{i}$ is assumed to be fully determined by the local environment around atom $i$. To represent the local environment, the distribution of other atoms around the atom $i$ is an obvious starting point. This distribution is represented by the probability density $\rho_{i}$ to find another atom $j$ at the position $\mathbf{r}$ around the atom $i$ within a radius $R_{\mathrm{cut}}$. It is defined as
\begin{align}
\rho_{i}\left(\mathbf{r}\right) &= \sum\limits_{j=1}^{N_{\mathrm{a}}} f_{\mathrm{cut}}\left(r_{ij}\right) g\left(\mathbf{r}-\mathbf{r}_{ij}\right), \label{equation_probability_density}
\end{align}
where $f_{\mathrm{cut}}$ is a cutoff function that smoothly removes the information outside the radius $R_{\mathrm{cut}}$. The position vector of the atom $i$ is denoted by $\mathbf{r}_{i}$, and $r_{ij}=|\mathbf{r}_{ij}|=|\mathbf{r}_{j}-\mathbf{r}_{i}|$ is the distance between two atoms, and $g(\mathbf{r})$ is the delta function $\delta(\mathbf{r})$. In SOAP, the delta function is replaced by a normalized Gaussian function as
\begin{align}
g\left(\mathbf{r}\right)&=\frac{1}{\sqrt{2\sigma_{\mathrm{atom}}\pi}}\mathrm{exp}\left(-\frac{|\mathbf{r}|^{2}}{2\sigma_{\mathrm{atom}}^{2}}\right). \label{equation_single_atom_distribution}
\end{align}
The local energies $U_{i}$ are functionals of the density $\rho_{i}$, $U_{i}$=$F[\rho_{i}(\mathbf{r})]$. So the simplest numerical approach to implement this procedure
would be to develop $\rho_{i}(\mathbf{r})$ into a finite basis set and express $F$ as a function of the coefficients. The drawback of this is that $F$ would not 
possess  rotational invariance. Therefore, it is expedient to introduce intermediate functions--- usually called {\em descriptors} ---that depend on $\rho_i(\mathbf{r})$. These intermediate
functions should be invariant under rotations (as well as translations). The simplest rotationally invariant descriptor is the radial distribution function defined as
\begin{align}
\rho_{i}^{(2)}\left(r\right) &= \frac{1}{4\pi} \int \rho_{i}\left(r\hat{\mathbf{r}}\right) d\hat{\mathbf{r}}, \label{equation_radial_distribution}
\end{align}
which measures pairwise distances from the atom $i$ within $R_\mathrm{cut}$ as schematically shown in Fig.~\ref{fig2} (a). Here, $\hat{\mathbf{r}}$ denotes the unit vector of $\mathbf{r}$. This function, however, cannot accurately describe the potential energy surface because of the lack of angular information. Specifically two different probability densities $\rho_i$ can yield an identical $\rho_{i}^{(2)}$, which
would then yield the same local energy $U_i$.
The necessary angular information can be incorporated by using the  probability to find an atom $j$ at a distance $r$ from the $i$th atom and another atom $k$ at a distance $s$ from the $i$th atom  along the angle $\angle{kij}=\theta$ as schematically shown in Fig.~\ref{fig2} (b). Starting from $\rho_{i}(\mathbf{r})$, this probability can be determined as
\begin{align}
\rho_{i}^{(3)}\left(r,s,\theta\right) &= \iint \delta \left(\hat{\mathbf{r}}\cdot\hat{\mathbf{s}} - \mathrm{cos}\theta\right) \rho_{i}\left(r\hat{\mathbf{r}}\right) \rho_{i}^{*} \left(s\hat{\mathbf{s}}\right) d\hat{\mathbf{r}} d\hat{\mathbf{s}}. \label{equation_angular_distribution}
\end{align}
This function, commonly referred to as angular distribution function, is equivalent to the power spectrum used in practical applications of the GAP~\cite{10,18,19,20}. In order to show this equivalence, $\rho_{i}$ is expanded as
\begin{align}
\rho_{i} \left( \mathbf{r} \right) &= \sum\limits_{l=1}^{L_{\mathrm{max}}} \sum\limits_{m=-l}^{l} \sum\limits_{n=1}^{N^{l}_{ \mathrm{R}}} c_{nlm}^{i}\chi_{nl} \left( r \right) Y_{lm} \left( \hat{\mathbf{r}} \right). \label{equation_density_expansion}
\end{align}
Here, $\big\{\chi_{nl}|n=1,...,N^{l}_{\mathrm{R}}, l=0,...,L_{\mathrm{max}}\big\}$ denote radial basis functions that satisfy the following orthonormal relation
\begin{align}
4\pi \int_{0}^{\infty} \chi_{nl} \left(r\right) \chi_{n'l} \left(r\right) r^{2} dr &= \delta \left(n-n'\right). \label{equation_orthonormal_radial}
\end{align}
$Y_{lm}$ are the spherical harmonics. By using Eq.~(\ref{equation_density_expansion}), Eqs.~(\ref{equation_radial_distribution}) and (\ref{equation_angular_distribution}) can be rewritten as
\begin{align}
\rho_{i}^{(2)}\left(r\right) &= \frac{1}{\sqrt{4\pi}} \sum\limits_{n=1}^{N^{0}_{\mathrm{R}}} c_{n}^{i} \chi_{nl}\left(r\right),  \\
c_{n}^{i}&= c_{n00}^{i}, \label{equation_cn} \\  
\rho_{i}^{(3)}\left(r,s,\theta\right) &= \sum\limits_{l=1}^{L_{\mathrm{max}}} \sum\limits_{n=1}^{N^{l}_{\mathrm{R}}}\sum\limits_{\nu=1}^{N^{l}_{\mathrm{R}}} \sqrt{\frac{2l+1}{2}} \nonumber \\
&\times p_{n\nu l}^{i}\chi_{nl}\left(r\right)\chi_{\nu l}\left(s\right)P_{l}\left(\mathrm{cos}\theta\right), \label{equation_angular_expansion} \\
p_{n\nu l}^{i}&=\sqrt{\frac{8\pi^{2}}{2l+1}} \sum\limits_{m=-l}^{l} c_{nlm}^{i} c_{\nu lm}^{i*}, \label{equation_pnnl}\end{align}
where $P_{l}$ is a Legendre polynomial of order $l$. Eq.~(\ref{equation_pnnl}) is the same as the equation for the power spectrum described in Refs.~[\onlinecite{22},\onlinecite{21}]. Eqs.~(\ref{equation_angular_expansion}) and (\ref{equation_pnnl}) indicate that $p_{n\nu l}^{i}$ corresponds to the expansion coefficients of $\rho_{i}^{(3)}$ with respect to the orthonormal radial and angular basis functions. Thus, $p_{n\nu l}^{i}$ contains the same information as the angular distribution defined in Eq.~(\ref{equation_angular_distribution}).
\subsection{Potential energy, Gaussian approximation potential }

In our ML algorithm, we use the distributions $\rho_{i}^{(2)}$ and $\rho_{i}^{(3)}$ to parameterize the potential energy surface  $U$. This means that $U_{i}$ is described as a functional of $\rho_{i}^{(2)}$ and $\rho_{i}^{(3)}$,
\begin{align}
U_{i}&=F\left[\rho_{i}^{(2)},\rho_{i}^{(3)}\right]. \label{equation_U_functional}
\end{align}
Obviously, it is not  generally a simple matter to find a suitable functional form, although neural networks and the moment tensor potentials (MTP) have been  found to yield an excellent approximation for total energies\cite{5,MTP1,MTP2}. 
In the present work we adopt the Gaussian approximation potential as pioneered by Bart\'ok and coworkers \cite{6}.
In this approach a set of $N_\mathrm{B}$ local reference structures $\{ \rho_{i_\mathrm{B}}| i_\mathrm{B}=1,...,N_\mathrm{B}\}$ are chosen. These reference
configurations are converted to a set of coefficients in the descriptor space $\{ \mathbf{X}_{i_\mathrm{B}}| i_\mathrm{B}=1,...,N_\mathrm{B}\}$
and the potential energy is approximated by fitting a set of coefficients $\{w_{i_\mathrm{B}}| i_\mathrm{B}=1,...,N_\mathrm{B}\}$:
\begin{align}
F\left[\rho_{i}^{(2)},\rho_{i}^{(3)}\right]&=\sum\limits_{i_\mathrm{B}=1}^{N_\mathrm{B}} w_{i_\mathrm{B}} K\left(\mathbf{X}_{i},\mathbf{X}_{i_\mathrm{B}}\right). \label{equation_U_smeard_expansion}
\end{align}
Here each  vector $\mathbf{X}_{i}$ collects all coefficients $c_{n}^{i}$ and $p_{n\nu l}^{i}$ for a specific local configuration $\rho_i(\mathbf{r})$ [Eqs.~(\ref{equation_cn}) and (\ref{equation_pnnl})]. 
The kernel $K$ is supposed to measure the similarity between a  local configuration of interest $\rho_i(\mathbf{r})$ and the reference configurations $\rho_{i_\mathrm{B}}(\mathbf{r})$. It usually approaches unity if two configurations are similar and decays towards a small value if the two configurations are different. 

In the present case, the following polynomial function is used
\begin{align}
K\left(\mathbf{X}_{i},\mathbf{X}_{i_\mathrm{B}}\right)&=\beta^{(2)}\left(\mathbf{X}_{i}^{(2)}\cdot\mathbf{X}_{i_{\mathrm{B}}}^{(2)}\right)+\beta^{(3)}\left(\mathbf{\bar{X}}_{i}^{(3)}\cdot\mathbf{\bar{X}}_{i_{\mathrm{B}}}^{(3)}\right)^{\zeta^{(3)}}. \label{equation_similarity_kernel}
\end{align}
Here, $\mathbf{X}_{i}^{(2)}$ and $\mathbf{X}_{i}^{(3)}$ are the vectors containing $c_{n}^{i}$ and $p_{n\nu l}^{i}$, respectively. The vector $\mathbf{\bar{X}}_{i}^{(3)}$ denotes a normalized vector of $\mathbf{X}_{i}^{(3)}$, $\beta^{(2)}$ and $\beta^{(3)}$ are weighting parameters, and $\zeta^{(3)}$ is a parameter to control the sharpness of the kernel. Phenomenologically, the first term in Eq.~(\ref{equation_similarity_kernel}) can be regarded as a pairwise linear interaction term, which is suited to describe long-range radial interactions, such as Coulomb and Lennard-Jones interactions. In contrast, the second term provides non-linear many-body interaction terms as discussed by Glielmo et al.~\cite{23}. The latter term is known as the SOAP~\cite{22}. The name SOAP relates to the fact that the dot product $\mathbf{X}_{i}^{(3)}\cdot\mathbf{X}_{i_{\mathrm{B}}}^{(3)}$ can be related to  the Haar integral~\cite{24} of the square of an overlap between two probability distributions as
\begin{align}
\mathbf{X}_{i}^{(3)}\cdot\mathbf{X}_{i_{\mathrm{B}}}^{(3)} &= \int |S_{ii_{\mathrm{B}}} (\hat{R})|^{2} d\hat{R}, \label{equation_Haar_integral} \\
S_{ii_{\mathrm{B}}} &= \int \rho_{i}\left(\mathbf{r}\right)\rho_{i_{\mathrm{B}}}(\hat{R}\mathbf{r}) d\mathbf{r}, \label{equation_overlap_integral3}
\end{align}
where $\hat{R}$ denotes the rotational operator, and the integral in (\ref{equation_Haar_integral}) needs to be performed over all possible rotations defined by this operator. In this sense, SOAP measures the structural similarity between the structure $i$ and the reference structure $i_{\mathrm{B}}$ as an overlap of two probability distributions. Similarly, the dot product $\mathbf{X}_{i}^{(2)}\cdot\mathbf{X}_{i_{\mathrm{B}}}^{(2)}$ can be rewritten as an overlap integral between the radial distribution functions as
\begin{align}
\mathbf{X}_{i}^{(2)}\cdot\mathbf{X}_{i_{\mathrm{B}}}^{(2)} &= 4\pi \int_{0}^{R_{\mathrm{cut}}}{\rho_{i}^{(2)}\left(r\right)\rho_{i_{\mathrm{B}}}^{(2)}\left(r\right)}r^{2}dr. \label{equation_overlap_integral2}
\end{align}
In this study, in order to examine the efficiency of the on-the-fly scheme in a manner comparable to previous publications using the  SOAP scheme, we only use the SOAP kernel  $K(\mathbf{X}_{i},\mathbf{X}_{i_\mathrm{B}})$ by setting $\beta^{(2)}=0$ for all materials. An application including the radial descriptor is presented elsewhere~\cite{25}.

Before ending this subsection, we briefly explain some key points where our implementation differs from the previous SOAP implementations. At every MD step, the expansion coefficients $c_{nlm}^{i}$ and their partial derivatives with respect to the atomic positions need to be calculated in order to compute the potential energy $U$ and its partial derivatives. In the SOAP~\cite{22}, $c_{nlm}^{i}$ can be analytically formulated using the Gaussian function of Eq.~(\ref{equation_single_atom_distribution}) as
\begin{align}
c_{nlm}^{i}&=\sum\limits_{j=1}^{N_{\mathrm{a}}} h_{nl}\left(r_{ij}\right) Y^{*}_{lm}\left(\hat{\mathbf{r}}_{ij}\right), \label{equation_expansion_coefficient} \\
h_{nl}\left(r\right)&=\frac{4\pi}{\left(\sqrt{2\sigma_\mathrm{atom}^{2}\pi}\right)^{3}} f_\mathrm{cut}\left(r\right)\int_{0}^{\infty} \chi_{nl}\left(r'\right) \nonumber \\
&\times\mathrm{exp}\left(-\frac{r'^{2}+r^{2}}{2\sigma_\mathrm{atom}^{2}}\right)\iota_{l}\left(\frac{rr'}{\sigma_{\mathrm{atom}}^{2}}\right)r'^{2}dr', \label{equation_radial_function}
\end{align}
where $\iota_{l}$ denotes the modified spherical Bessel function of the first kind. However, the calculation of $h_{nl}$ and $dh_{nl}/dr$ by Eq.~(\ref{equation_radial_function}) would be computationally rather demanding. In order to accelerate the calculations, we adopt a spline interpolation. In this method, $h_{nl}\left(r\right)$ is calculated on a radial mesh over $r$ once at the beginning of the training or MD simulation. The calculated function is spline-interpolated, and the interpolated function is used to calculate the coefficients $c^{i}_{nlm}$ and their derivatives later. Another difference to previous implementations is in the choice of the radial basis functions. In our method, normalized spherical Bessel functions $\chi_{nl}= j_{l}\left(q_{nl}r\right)$ are used as the radial basis functions, because their mutual orthogonality allows for a systematic improvement by simply increasing the number $N^{l}_{\mathrm{R}}$ of the radial functions as described in Appendix A. Finally, we use the cutoff function proposed by Behler and Parrinello~\cite{5} defined as
\begin{align}
f_{\mathrm{cut}}\left(r_{ij}\right) & = \begin{cases} \frac{1}{2} \left[ \mathrm{cos}\left(\pi \frac{r_{ij}}{R_{\mathrm{cut}}}\right)+1 \right] & \mbox{if } r_{ij} \leq R_{\mathrm{cut}} \\ 0, & \mbox{otherwise} \end{cases}. \label{equation_fcut}
\end{align}
This cutoff function weights atoms close to the central atom $i$  more strongly. This radially scaled weight enables the descriptor to efficiently describe the structural differences that strongly influence the potential energy of atom $i$ similarly to the radially scaled kernel proposed by Willat and co-workers~\cite{56}.
In Table S1 in the Supplemental Material (SM)~\cite{SM} all parameters described in the subsections A and B are tabulated.
\subsection{Fitting of energy, forces and stress tensor and their uncertainty}
In order to determine the fitting parameters $w_{i_{\mathrm{B}}}$, the energies, forces and stress tensor for a
set of reference structural datasets labeled by a superscript $\alpha=1,...,N_\mathrm{st}$ must be fitted. 
Combining Eqs~(\ref{equation_pes}), (\ref{equation_U_functional}) 
and (\ref{equation_U_smeard_expansion}) yields for the total energy per atom the following equation 
that must be fulfilled in a least square sense:
\begin{align}
\frac{U^{\alpha}}{N^{\alpha}_{\mathrm{a}}} &\overset{!}{=} \sum\limits_{i=1}^{N_{\mathrm{a}}^{\alpha}} \frac{U_{i}^{\alpha}}{N^{\alpha}_{\mathrm{a}}} \nonumber \\
                           &= \sum\limits_{i_\mathrm{B}=1}^{N_\mathrm{B}} w_{i_\mathrm{B}} \sum\limits_{i=1}^{N_{\mathrm{a}}^{\alpha}} \frac{K\left(\mathbf{X}_{i}^{\alpha},\mathbf{X}_{i_\mathrm{B}}\right)}{N^{\alpha}_{\mathrm{a}}} \quad \forall \alpha=1,...,N_\mathrm{st}. 
\end{align}
Here $U_{i}^{\alpha}$ is the local energy of atom $i$ in the structure $\alpha$, and $\mathbf{X}_{i}^{\alpha}$ 
is the vector of coefficients in the descriptor space for atom $i$ in structure $\alpha$, and $U^{\alpha}$ is the actual FP energy. 
In practice, we simultaneously fit the energy per atom, 
forces and the stress tensor for reference structures $\alpha$, for which FP calculations have yet been performed (see below for details). 
The previous equation indicates that the total potential energy $U^{\alpha}$ is a linear function of the coefficients $w_{i_{\mathrm{B}}}$. 
It is also straightforward to see that the forces and the stress tensor components are described as linear functions of the 
coefficients $w_{i_{\mathrm{B}}}$. These linear equations can be collected into a matrix-vector form as
\begin{align}
\mathbf{y}^{\alpha}& \overset{!}{=}\bm{\phi}^{\alpha}\mathbf{w} \quad \forall \alpha. \label{equation_linear_equation}
\end{align}
Here, $\{\mathbf{y}^{\alpha}|\alpha=1,...,N_\mathrm{st}\}$ denotes  column vectors containing 
the dimensionless FP potential energy per atom in the first line, the forces and the 
components of the stress tensor in the subsequent lines for 
a single structure $\alpha$ in the reference structure dataset, in total 
\begin{equation}
m^{\alpha}=1+3 N_{\mathrm{a}}^{\alpha} +6
\label{equ:m}
\end{equation} 
components for $N_{\mathrm{a}}^{\alpha}$ atoms. The entries are made dimensionless by dividing their 
values by the standard deviations of the FP energies per atom, forces and stress tensors in the reference structure datasets. 
The column vector $\mathbf{w}$ is comprised of the $N_{\mathrm{B}}$ coefficients $w_{i_{\mathrm{B}}}$, 
and $\bm{\phi}^{\alpha}$ is a $m^{\alpha} \times N_{\mathrm{B}}$ matrix. The first line of the matrix is made up by 
$\sum_i K\left(\mathbf{X}^{\alpha}_{i},\mathbf{X}_{i_\mathrm{B}}\right)/ N^{\alpha}_{\mathrm{a}}$, the second line is made up by the derivative 
of the energy with respect to the first atomic coordinate in the structure $\alpha$ and so on. 

After the fitting, the energies and forces of a new structure with the descriptor $\mathbf{X}_{i}$ can be efficiently obtained calculating
\begin{align}
\mathbf{y} &=\bm{\phi} \mathbf{w}, \label{equation_prediction_y}
\end{align}
where $\bm{\phi}$ comprises $\sum_i K\left(\mathbf{X}_{i},\mathbf{X}_{i_\mathrm{B}}\right)$ in the first row,
and the partial derivatives of the kernel with respect to the coordinates in the structure in the subsequent rows.

In the conventional schemes, FP calculations are carried out on a wide variety of structures in advance, 
local configurations are chosen from each structure in the dataset, and the coefficients $\mathbf{w}$ 
are optimized to optimally reproduce the reference structure datasets. In our scheme,
the FP data generation, selection, and parameter optimization are carried out on the fly during the MD simulations.
A key component that makes this algorithm extremely efficient is the evaluation of the uncertainty in the prediction.
This is used to decide whether the FP calculations are necessary or not at step 2).
In our scheme, both the optimization of $\mathbf{w}$ and the uncertainties are 
estimated by a Bayesian linear-regression method~\cite{26}.

The Bayesian linear regression assumes the presence of a set of the coefficients that exactly reproduce
the exact energy, forces and stress tensor without any numerical noises. Then, this method determines the probability 
to find this exact set of coefficients at $\mathbf{w}$ on the basis of the observation of a limited number of FP datasets
$\{\mathbf{y}^{\alpha}|\alpha=1,...,N_\mathrm{st}\}$ containing numerical noises. In practice, the FP data
usually carry comparatively small errors, however, the finite cutoff $R_{\mathrm{cut}}$
implies that the model can never exactly describe the FP data. In other words,
as the atoms outside the cutoff radius move, the local energies and forces should change, but
since the model assumes that the local energy depends only  on the position of the atoms
inside the cutoff sphere, residual errors are introduced. We assume
that these errors can be modeled by the presence of noise with a Gaussian distribution  in the FP data.
This assumption also implies some errors in the model parameters  $\mathbf{w}$.
The so-called posterior distribution
$p\left( \mathbf{w} | \mathbf{Y} \right)$ is determined as a Gaussian distribution 
written as (see necessary assumptions in Appendix B):
\begin{align}
p\left( \mathbf{w} | \mathbf{Y} \right) &= \mathcal{N} \left( \mathbf{\bar w},\mathbf{\Sigma} \right), \label{equation_posterior_distribution_w} \\
\mathbf{\bar w} &= \frac{1}{\sigma_{\mathrm{v}}^{2}} \mathbf{\Sigma} \mathbf{\Phi}^{\mathrm{T}} \mathbf{Y}, \label{equation_wm} \\
\mathbf{\Sigma}^{-1} &=\frac{1}{\sigma_{\mathrm{w}}^{2}} \mathbf{I} + \frac{1}{\sigma_{\mathrm{v}}^{2}} \mathbf{\Phi}^{\mathrm{T}}\mathbf{\Phi}. \label{equation_Sigma}
\end{align}
Here, ${\mathbf{Y}}$ is a super vector with size $M=\sum_{\alpha} m^{\alpha}$ [compare Eq.~(\ref{equ:m})] collecting {\em all} FP energies per atom, forces and stress tensors
$\{\mathbf{y}^{\alpha}|\alpha=1,...,N_\mathrm{st}\}$ in the reference structure datasets. 
Similarly, the $M \times N_{\mathrm{B}}$ design matrix $\mathbf{\Phi}$ is a collection of
all matrices $\bm{\phi}^{\alpha}$ on all reference structure datasets, and $\mathbf{I}$ denotes the unit matrix.
The symbols $\sigma_{\mathrm{v}}^{2}$ and $\sigma_{\mathrm{w}}^{2}$ denote parameters optimized to balance the accuracy 
and robustness of the evolving force field as explained later on.
The symbol $\mathcal{N} \left( \mathbf{\bar w},\mathbf{\Sigma} \right)$ is a multidimensional normalized Gaussian centered at $ \mathbf{\bar w}$ and
defined as
\begin{align}
\mathcal{N} \left( \mathbf{\bar w},\mathbf{\Sigma} \right) &= \frac{1}{\sqrt{ \left( 2 \pi \right)^{N_{\mathrm{B}}} ||\bm {\Sigma}|| }} \nonumber \\
& \times \mathrm{exp} \left[ -\frac{ \left( \mathbf{w} - \mathbf{\bar w} \right)^{\mathrm{T}} \bm{\Sigma}^{-1} \left( \mathbf{w} - \mathbf{ \bar w} \right)}{2} \right], \label{equation_definition_N}
\end{align}
where $||\bm{\Sigma}||$ means the determinant of the matrix $\bm{\Sigma}$. The desired optimal coefficients are determined at the center of the Gaussian distribution $\mathbf{w}=\mathbf{\bar w}$, 
where the posterior probability is maximized. It is straightforward to show that the vector $\mathbf{\bar w}$ is  identical to the vector obtained by the ridge regression, with the ratio of  $\sigma_{\mathrm{v}}^{2}$ to $\sigma_{\mathrm{w}}^{2}$ being equivalent to the Tikhonov regularization parameter.

The uncertainty in the predictions is provided from the probability to find the exact FP energy per atom, forces and stress tensor 
at $\mathbf{y}$. This posterior distribution $p \left( \mathbf{y} | \mathbf{Y} \right)$ is obtained from the posterior distribution 
$p\left( \mathbf{w} | \mathbf{Y} \right)$ as (see also Appendix B)
\begin{align}
p \left( \mathbf{y} | \mathbf{Y} \right) &= \mathcal{N} \left( \bm{\phi}\mathbf{\bar w}, \bm{\sigma} \right), \label{equation_posterior_distribution_y} \\
\bm{\sigma}&=\sigma_{\mathrm{v}}^{2}\mathbf{I}+\bm{\phi}^{\mathrm{T}}\mathbf{\Sigma}\bm{\phi}. \label{equation_sigma}
\end{align}
The mean vector $\bm{\phi}\mathbf{\bar w}$ contains the results of the predictions on the dimensionless energy per atom, forces and stress tensor. 
The diagonal elements of $\bm{\sigma}$, which correspond to the variances of the predicted results, are used as the uncertainty in the prediction.

As in the ridge regression, the optimization of the parameters $\sigma_{\mathrm{v}}^{2}$ and $\sigma_{\mathrm{w}}^{2}$ is
important to prevent overfitting. In our algorithm, 
they are optimized by the evidence approximation~\cite{27,28,29}. 
In this scheme, the parameters $\sigma_{\mathrm{v}}^{2}$ and $\sigma_{\mathrm{w}}^{2}$ are determined by 
maximizing the marginal likelihood function called evidence function. This evidence function corresponds 
to a probability that the regression model with specific parameters $\sigma_{\mathrm{v}}^{2}$ and 
$\sigma_{\mathrm{w}}^{2}$ provides the reference data $\mathbf{Y}$, and it is calculated as 
\begin{align}
p\left(\mathbf{Y}|\sigma_{\mathrm{v}}^{2},\sigma_{\mathrm{w}}^{2}\right)&=\left(\frac{1}{\sqrt{2\pi\sigma_{\mathrm{v}}^{2}}}\right)^{M} \left(\frac{1}{\sqrt{2\pi\sigma_{\mathrm{w}}^{2}}}\right)^{N_{\mathrm{B}}} \nonumber \\
&\times\int \mathrm{exp}\left[-E\left( \mathbf{w} \right) \right] d\mathbf{w}, \label{equation_likelihood} \\
E\left( \mathbf{w} \right) &= \frac{1}{2\sigma_{\mathrm{v}}^{2}} || \mathbf{\Phi}\mathbf{w}-\mathbf{Y}||^{2} + \frac{1}{ 2\sigma_{\mathrm{w}}^{2}}||\mathbf{w}||^{2}. \label{equation_loss}
\end{align}
Hence, the optimization over $\sigma_{\mathrm{v}}^{2}$ and $\sigma_{\mathrm{w}}^{2}$ can be regarded 
as the maximization of the probability to provide the correct answer $\mathbf{Y}$ at any regression 
coefficient $\mathbf{w}$. Details of the maximization method are documented in Appendix C.
\begin{figure}
\includegraphics[width=0.40\textwidth,angle=0]{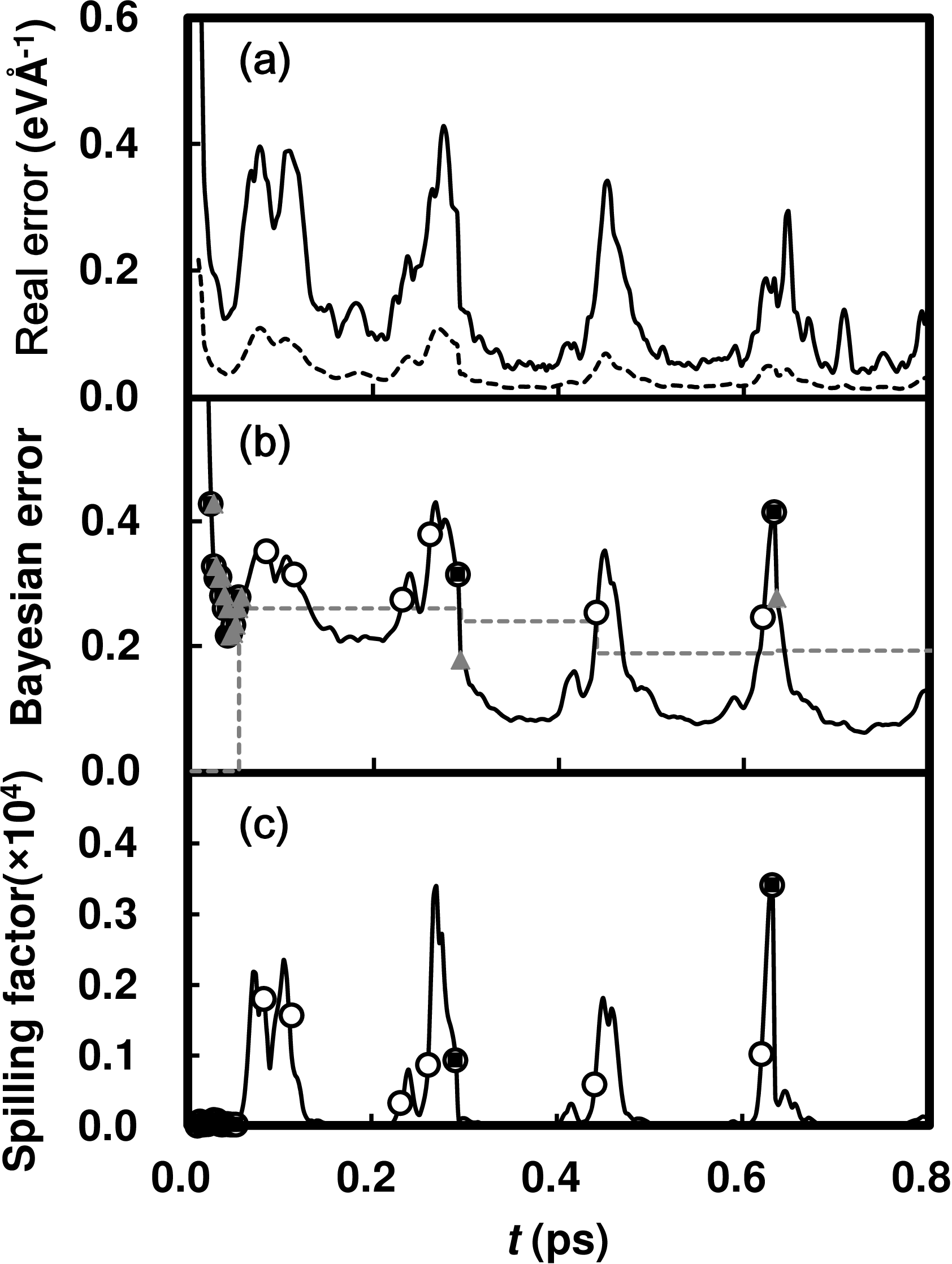}
\caption{Errors during a finite temperature simulation of rocksalt  MgO at 300 K. (a) Actual difference between the FP forces and the forces calculated by the MLFF. Solid line shows  the maximum difference on a single atom, dotted line shows the root mean square difference averaged over all atoms. (b) The maximum Bayesian error of the  dimensionless force on a single atom calculated by Eq.~(\ref{equation_sigma}) and scaled to the real error. 
(c) The maximum spilling factor on a single atom. Open circles and black squares indicate step 3) and step 4), respectively. Dark gray triangles indicate the Bayesian errors stored as $\sigma_{\mathrm{max},I}$. These are used to determine the threshold for the Bayesian error $\epsilon_{\mathrm{BE}}$ as explained in Section E. Gray dashed lines indicate the threshold for the Bayesian error.}
\label{fig3}
\end{figure}

In addition to the Bayesian error, we also calculate the spilling factor suggested by Miwa and Ohno~\cite{30}. 
The spilling factor is a measure of the density of local reference configurations $i_{\mathrm{B}}$ near 
the new local configuration ${\mathbf{X}}$ in the descriptor space, and its equation is formulated as
\begin{align}
s &= 1 \nonumber \\
&- \frac{ \sum\limits_{i_{\mathrm{B}}=1}^{N_{\mathrm{B}}} \sum\limits_{i'_{\mathrm{B}}=1}^{N_{\mathrm{B}}} K\left(\mathbf{X},\mathbf{X}_{i_{\mathrm{B}}}\right) K^{-1}\left(\mathbf{X}_{i_{\mathrm{B}}},\mathbf{X}_{i'_{\mathrm{B}}}\right) K\left(\mathbf{X}_{i'_{\mathrm{B}}},\mathbf{X}\right)}{K\left(\mathbf{X},\mathbf{X}\right)}. \label{equation_sf}
\end{align}
It should be noted that the equation is slightly modified from its original equation in order to make 
it applicable to a non-normalized similarity measure such as the one used by us, see 
Eq.~(\ref{equation_similarity_kernel}). If the density of the reference configuration is great enough to 
provide complete overlap among the configurations, $s$ approaches zero, otherwise $s$ approaches unity.

Trends of these two estimated errors are illustrated in Fig.~\ref{fig3}, 
where the real errors, Bayesian errors and spilling factors are shown as a function of 
the MD simulation time during on-the-fly force field generations for the MgO solid at 300 K. 
The Bayesian error estimation can correctly detect the large errors at the beginning of 
the training because of the presence of the regularization parameters $\sigma_{\mathrm{v}}^{2}$ and $\sigma_{\mathrm{w}}^{2}$,
which can describe the uncertainty in the coefficients $\mathbf{w}$ caused by too few reference data. 
The regularization parameters also stabilize the computations of the covariance matrix $\mathbf{\Sigma}$. 
These parameters, however, make the Bayesian error less sensitive to structural differences. 
In contrast, because of the lack of these regularization parameters, the spilling factor can more sensitively 
detect the uncertainty caused by the structural difference. However, numerical instabilities can 
occur in its computation because the matrix $K$ is not regularized. This problem is critical particularly 
in simple materials examined in this study, where the diversity is small in the appearing local configurations, 
and the spilling factor is less than $10^{-4}$ in most cases because of strong overlap 
among the selected local reference configurations. As indicated in Fig.~\ref{fig3}, the small spilling factor 
does not mean that the data sampling is unnecessary. The real error can be relatively large, 
and the spilling factor captures only the qualitative trends of the real error. 
Thus, the machine needs to make the decision at step 2) on the basis of the unreliable small 
spilling factor interfered by the numerical instability.

Details of the decision scheme at step 2) are explained in the following subsection. 
Here we briefly note that the criterion for the spilling factor is set to a relatively large value of 
0.02 as in~Ref.[\onlinecite{26}]. For the applications reported here,  the threshold for the spilling factor is hardly
ever passed. Only for liquid and interfacial MgO, 
the calculated spilling factors exceeded this criterion. But even then only 2 \% of the 
total FP calculations are executed because of the spilling factor criterion.
So in the present work, the Bayesian error criterion is more relevant.

\begin{figure}
\includegraphics[width=0.40\textwidth,angle=0]{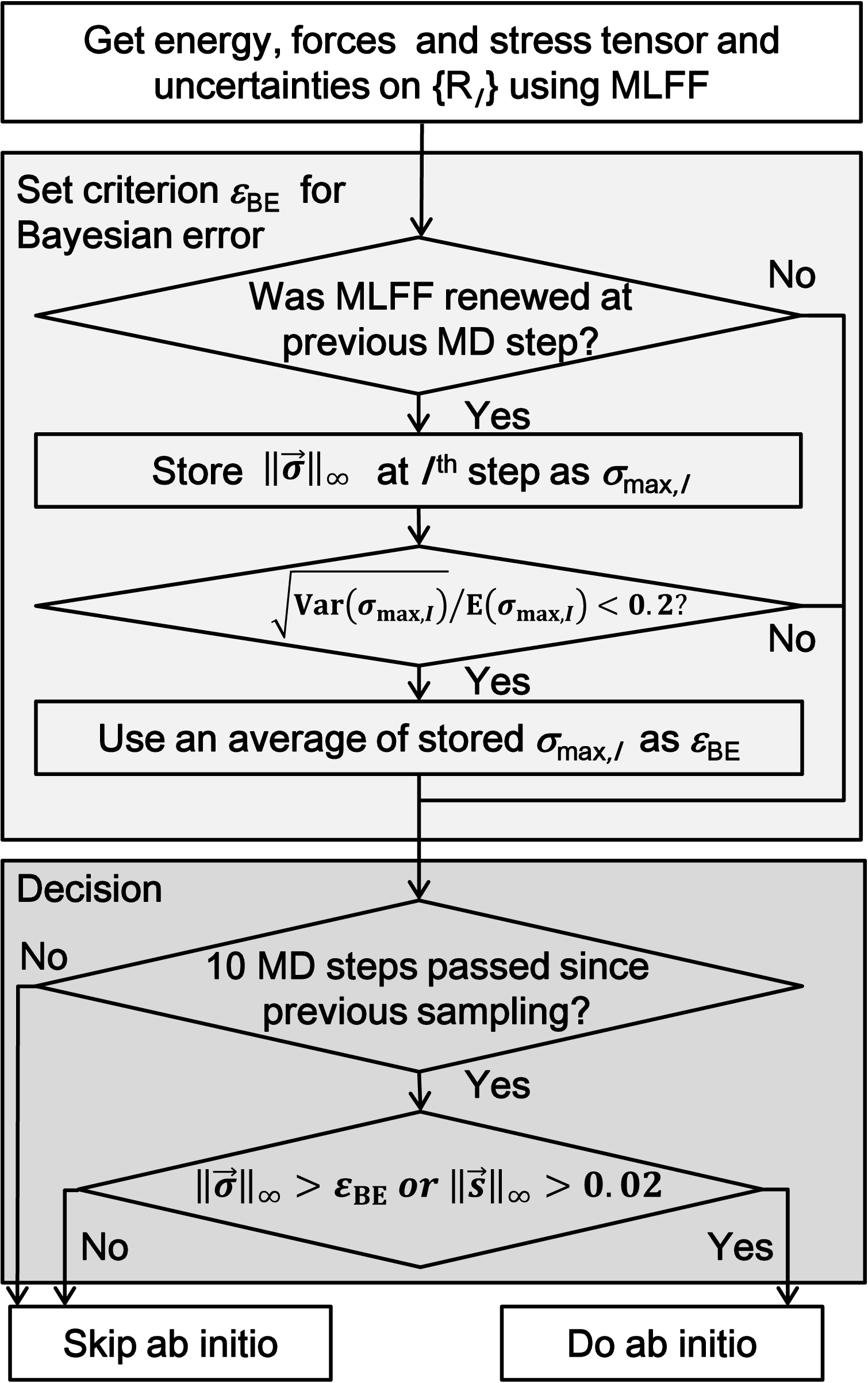}
\caption{Flowchart of the decision step whether to perform FP simulations or not. The symbol $\vec{\sigma}$ and $\vec{s}$ denote the vectors containing the Bayesian errors in the forces and spilling factors, respectively, for all atoms.  $||\vec{x}||_{\infty}$ denotes the infinity norm (also called supremum norm) of the vector $\vec{x}$, $\epsilon_{\mathrm{BE}}$ denotes the criterion for the Bayesian error, and $\mathrm{Var}\left(x\right)$ and $\mathrm{E}\left(x\right)$ refer to the  the variance and the average of the data $x$. At the beginning of every training simulation, the criterion $\epsilon_{\mathrm{BE}}$ is set to zero. }
\label{fig4}
\end{figure}
\subsection{Decision to perform FP calculation}
The decision whether to perform a FP calculation in step 2) or not 
(see section~\ref{subsection_outline_of_on_the_fly}) is obviously an important one. 
Figure~\ref{fig4} shows the flowchart of our decision scheme. As shown in the dark gray square 
in this figure, the decision is done on the basis of the estimated errors and 
the history of the previous samplings. First, the machine checks the previous data sampling step. 
If the current step is within 10 MD steps from the previous sampling step, 
the machine skips the FP calculations. This process avoids too dense sampling within a narrow phase space. 
If more than 10 MD steps have passed since the previous sampling, the machine examines the estimated errors. 
If the maximum Bayesian error in one of the forces or the spilling factor is larger than the chosen threshold, 
the machine performs the FP calculation, otherwise the FP calculation is skipped. 

The threshold for the spilling factor is set to 0.02 following Ref.~[\onlinecite{30}].
For the Bayesian error, which exhibits a descriptor- and materials-dependent non-zero value, 
the threshold $\epsilon_{\mathrm{BE}}$ is automatically determined on the fly. 
The corresponding scheme is shown in the light gray square in Fig.~\ref{fig4}. 
At the MD step $I$ just after the refinement of the force field (shown as the gray triangles in Fig.~\ref{fig3}), 
the machine stores the maximum value $\sigma_{\mathrm{max},I}$ of the Bayesian errors of 
the forces predicted for the new structure $I$. Because this new structure does not significantly 
differ from the structure sampled at the training step, the calculated Bayesian errors are 
nearly identical to the Bayesian errors on the previously sampled structure.
Hence, this maximum Bayesian error $\sigma_{\mathrm{max},I}$ can be regarded as a 
measure of the lowest currently attainable Bayesian error and can provide a reasonable threshold 
for the future. In our algorithm, the threshold is updated to be the average of 
the last 10 $\sigma_{\mathrm{max},I}$, if their relative standard deviation is smaller than 0.2 (empirically set). 
The dashed gray lines in Fig.~\ref{fig3} (b) illustrate the criteria determined by this on-the-fly 
scheme during the training on the MgO solid as an example.

\begin{figure}
\includegraphics[width=0.40\textwidth,angle=0]{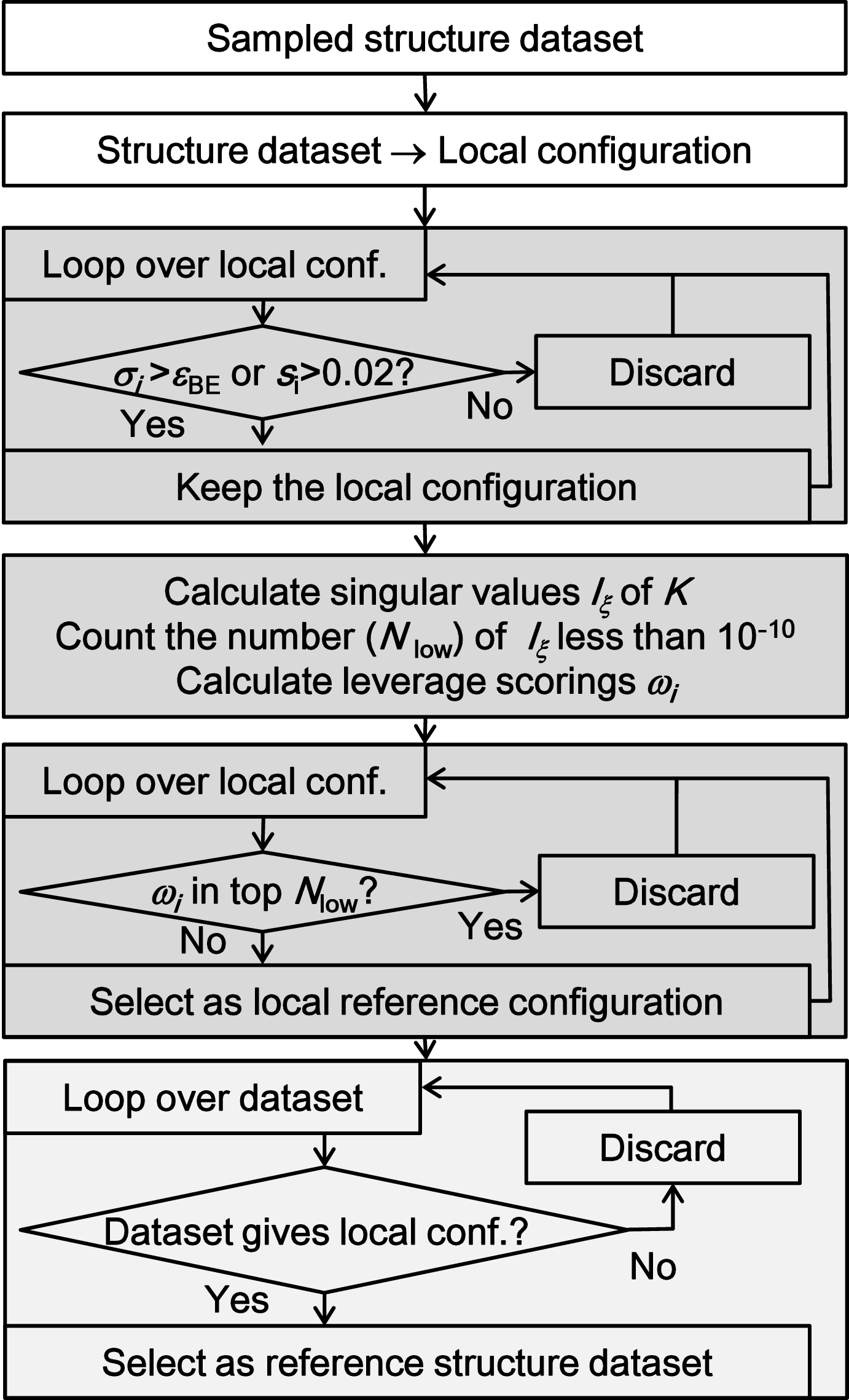}
\caption{Flowchart of the sparsification and data selection step. The symbol $K$ denotes the matrix comprised of the elements $K(i,j)$ defined as Eq.~(\ref{equation_similarity_kernel}). The leverage scoring $\omega_{i}$ is calculated by Eq.~(\ref{equation_ls1}) in Appendix D.}
\label{fig5}
\end{figure}
\subsection{Sparsification and data selection}
As previously explained, whenever the machine decides that for a specific structure insufficient information is stored in the machine, FP calculations are performed for that structure. To reduce the computational demands, the machine is not retrained after each FP calculation, but instead retraining is done typically after $n$=5 FP calculations or when the estimated errors are twice larger than the determined criteria. This allows to block many of the computationally expensive steps in the training. When $n$ FP calculations have been performed, the reference structure datasets and local reference configurations are selected, and the force field is refined. Figure~\ref{fig5} shows the flowchart corresponding to this data selection. As shown in the dark gray square, the data selection is done in  a two step procedure. First, the machine selects those local configurations that exhibit Bayesian errors on forces and spilling factors that are larger than the threshold. Although this step is already a sparsification process, numerical instabilities sometimes occur because of the overcompleteness among the remaining local configurations. In order to avoid those numerical instabilities, another sparsification process is performed using a CUR algorithm~\cite{31,19}. In our implementation, the machine examines correlations between the local configurations and eigenvalues of the matrix $K$  smaller than $\mathrm{10^{-10}}$ using leverage scoring $\omega_{i}$. For
details we refer to  Appendix D. Starting from the largest $\omega_{i}$ in descending order, $N_{\mathrm{low}}$ configurations are discarded, where $N_{\mathrm{low}}$ denotes the number of the eigenvalues smaller than $\mathrm{10^{-10}}$. Finally, as shown in the light gray square in Fig.~\ref{fig5}, the machine discards those structure datasets that do not provide any local reference configurations to speed up the computations and to reduce the memory usage.

\section{Technical details of melting point calculations}
\label{section_method_Tm}

\subsection{Training conditions}
For each material, the local reference configurations and the datasets were collected 
during the MD simulations on solid, liquid and interfacial systems. The Al solid and liquid 
were modeled by unit cells with 108 atoms, and the Al interface was modeled by a unit cell with 
144 atoms. For the other materials, solids and liquids were modeled by unit cells with 64 atoms, and 
the interfaces were modeled by unit cells with 128 atoms. A $3\times3\times3$ $\mathbf{k}$-point mesh 
was used for the Al solid and liquid, and a $3\times3\times2$ $\mathbf{k}$-point mesh 
was used for the Al interface. For the other materials, a $2\times2\times2$ mesh was used for the solids and liquids, and 
a $2\times2\times1$ mesh was used for the interfaces. Plane-wave basis sets and the projector augmented wave (PAW) method   
were used in all FP calculations. The PAW atomic reference configuration was $2\mathrm{s}^{2}2\mathrm{p}^{4}$ for O, 
$3\mathrm{s}^{2}3\mathrm{p}^{0}$ for Mg, $3\mathrm{s}^{2}3\mathrm{p}^{1}$ for Al, $3\mathrm{s}^{2}3\mathrm{p}^{2}$ for Si, 
$4\mathrm{s}^{2}4\mathrm{p}^{2}$ for Ge, and $5\mathrm{s}^{2}5\mathrm{p}^{2}$ for Sn.
The plane-wave cutoff energy was set to 325, 325, 225, 135 and 520 eV for Al, Si, Ge, Sn and MgO, respectively.
Training was performed for several functionals: the local density
approximation (LDA) in the parametrization of Ceperly and Alder~\cite{CA},
the Perdew-Burke-Ernzerhof (PBE) functional~\cite{PBE}, its variant for solids (PBEsol)~\cite{PBEsol}
and the strongly constrained appropriately normed (SCAN) functional~\cite{SCAN}.
For each material, the local reference configurations and the datasets were collected during the MD simulations on solid, liquid 
and interfacial systems. For each condition within all phases, the MD simulation was 
executed for 100 ps. The MD time step was set to 3 fs for all materials except for Ge and Sn, where the time step was set to 10 fs.
Further details of the MD parameters and snapshots  of the models are shown in Section S2 in the SM~\cite{SM}.

\begin{figure}
\includegraphics[width=0.40\textwidth,angle=0]{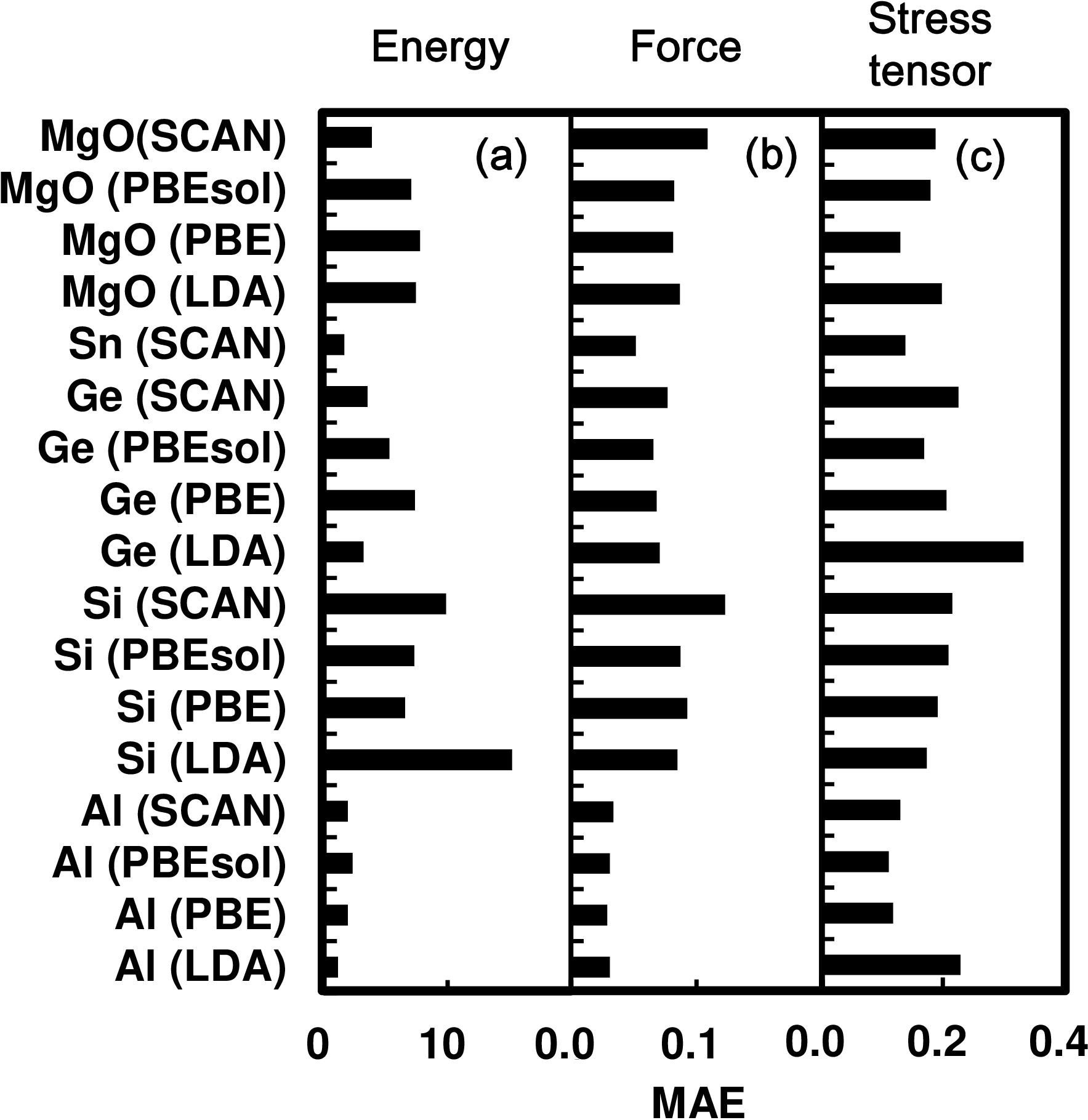}
\caption{Mean absolute errors in the energy per atom ($\mathrm{meV} \mathrm{atom}^{-1}$) (a), force ($\mathrm{eV} \mathrm{\AA}^{-1}$) and stress tensor (GPa) predicted by the force fields on 200 configurations of solids and liquids at the melting points.}
\label{fig6}
\end{figure}
\subsection{Efficiency and accuracy}
During the on-the-fly force field generation, more than 99 \% of the FP calculations were skipped reducing the computational time by a factor of more than 200. These accelerations allow the machine to efficiently collect reference configurations in a wide phase space. The details of the skipping ratio and the acceleration are summarized in Table S3 in the SM~\cite{SM}.

The number of structures in the reference structure datasets is typically less than 500, and the number of local reference configurations is less than 1000 as shown in Table S4 in the SM~\cite{SM}. Both are much smaller than the reference configurations used in previous studies~\cite{5,9,10,11,12,13}. It should be mentioned though that we train
our force fields essentially to the specific application, here liquids, solids and interfaces.

By the efficient sampling of the local reference configurations, MD simulations by the force fields are noticeably accelerated. Table S5 in the SM~\cite{SM} tabulates the elapsed time per MD step by the force fields and FP calculations. The force fields accelerate the MD simulations by factors of 2000 to 5000.

In addition to the significant acceleration of the computational speed, the adaptable reference datasets and flexible functional form allows for accurate predictions of the potential energy surfaces. The error analysis summarized in Fig.~\ref{fig6} indicates that the  mean absolute errors (in parentheses root mean square errors) in energies, forces and stress tensors are 5.5 (6.2) $\mathrm{meV} \mathrm{atom}^{-1}$, 0.07 (0.09) $\mathrm{eV} \mathrm{\AA}^{-1}$ and 0.18 (0.27) GPa, on average, respectively. 

\subsection{Interface pinning}
\begin{table*}
\caption{Melting temperatures (K) of Al, Si, Ge, Sn and MgO. CORR and MLFF denote the results with and without the thermodynamic perturbation corrections, respectively. CORR-$3^3$ denotes the results for Si, Ge and Sn with the thermodynamic perturbation using a $3\times 3\times 3$ $\mathbf{k}$-point mesh. Values in the parentheses indicate the uncertainties estimated by the block averaging method described in Refs.~[\onlinecite{54,55}].
Reference calculations with tight error tolerance  and similar  PAW potential are underlined.}
\label{table_1}
\begin{center}
\begin{tabular}{lcrrrrr|lcrrrr}
\hline
&\multicolumn{1}{c}{XC} &\multicolumn{1}{c}{MLFF} &\multicolumn{1}{c}{CORR} &\multicolumn{1}{c}{CORR-$3^3$} &\multicolumn{1}{c}{DFT} &\multicolumn{1}{c}{Exp} &  &\multicolumn{1}{c}{XC} &\multicolumn{1}{c}{MLFF} &\multicolumn{1}{c}{CORR} &\multicolumn{1}{c}{DFT} &\multicolumn{1}{c}{Exp}   \\
\hline
Si &LDA    &$1207(~5)$ &$1298(~5)$  &$1283(~5)$  &$1350(100)^{a}$             &$1685(2)^{b}$ &Al  &LDA    &$918(~7)$   &$909(~8)$   &$890(20)^{c}$               &$933.47^{d}$     \\
   &       &           &            &            &$1300(~50)^{e}$             &              &    &       &            &            &                            &                 \\
   &       &           &            &            &$1241(~20)^{f}$             &              &    &       &            &            &                            &                 \\
   &PBE    &$1409(~6)$ &$1431(~9)$  &$1450(~9)$  &$\underline{1449}(~10)^{g}$ &              &    &PBE    &$871(~8)$   &$837(~9)$   &                            &                 \\
   &PBEsol &$1145(10)$ &$1172(21)$  &$1213(21)$  &                            &              &    &PBEsol &$986(~5)$   &$954(~6)$   &$985(30)^{f}$               &                 \\
   &SCAN   &$1786(~7)$ &$1825(~7)$  &$1833(~7)$  &$\underline{1842}(~10)^{g}$ &              &    &SCAN   &$1017(12)$  &$981(16)$   &                            &                 \\
\hline
Ge &LDA    &$814(~7)$  &$814(~9)$   &$841(~9)$   &                            &$1210.4^{b}$  &MgO &LDA    &$3165(20)$  &$3243(21)$     &                            &$3040(100)^{h}$ \\
   &PBE    &$843(~5)$  &$876(~5)$   &$893(~5)$   &                            &              &    &PBE    &$2652(20)$  &$2698(23)$ &$\underline{2747}(34)^{i}$   &$3250(~25)^{j}$  \\
   &PBEsol &$758(~8)$  &$759(~8)$   &$792(~8)$   &                            &              &    &PBEsol &$2916(19)$  &$2981(20)$     &                            &                 \\
   &SCAN   &$1060(~2)$ &$1065(~3)$  &$1081(~3)$  &                            &              &    &SCAN   &$3079(23)$  &$3072(25)$ &$\underline{3032}(54)^{i}$   &                 \\
\hline
Sn &SCAN   &$468(11)$  &$459(13)$   &$459(13)$   &                            &$505^{k}$     &    &       &          &              &                            &                 \\
\hline
\multicolumn{12}{l}{\footnotesize{$^a$ Data from Ref.~[\onlinecite{1}].  $\: \: \qquad \qquad$ $^b$ Data from Ref.~[\onlinecite{36}]. $\: \qquad \qquad$ $^c$ Data from Ref.~[\onlinecite{2}].  $\: \: \qquad \qquad$  $^d$ Data from Ref.~[\onlinecite{37}].}} \\
\multicolumn{12}{l}{\footnotesize{$^e$ Data from Ref.~[\onlinecite{3}].  $\: \: \qquad \qquad$ $^f$ Data from Ref.~[\onlinecite{4}].}} \\
\multicolumn{12}{l}{\footnotesize{$^g$ Data from Ref.~[\onlinecite{34}] using the same PAW and $3\times 3\times 3\times$ $\mathbf{k}$-points.  $\: \qquad \qquad \qquad \qquad \qquad \qquad$ $^h$ Data from Ref.~[\onlinecite{39}].}} \\
\multicolumn{12}{l}{\footnotesize{$^i$ Data from Ref.~[\onlinecite{35}] using $2\mathrm{p}^{6}3\mathrm{s}^{2}$  PAWs for Mg.  $\: \: \qquad \qquad \qquad$ $^j$ Data from Ref.~[\onlinecite{40}].  $\qquad \qquad$ $^k$ Data from Ref.~[\onlinecite{38}]}}
\end{tabular}
\end{center}
\end{table*}

The melting-point calculations are carried out using force fields and using the interface pinning method~\cite{4}. This method has
shown to be able to accurately predict the melting temperature and pressure\cite{4}. In the interface pinning method, an MD simulation with constant temperature and pressure~\cite{32,33} is carried out on a solid-liquid interface. During the MD simulation, a harmonic bias potential is added to the potential energy $U$ in order to constrain the order parameter $Q$ of the system to an intermediate order parameter $a$ between the solid and the liquid phases as
\begin{align}
U'&=U+\frac{\kappa}{2}\left(Q-a\right)^{2}, \label{equation_bias_potential}
\end{align}
where $\kappa$ is a force constant. From the mean force that keeps the order parameter close to $a$, the difference in the chemical potential $\Delta\mu$ between two phases is calculated as
\begin{align}
\Delta\mu&=-\kappa\left(\langle Q \rangle'-a\right)\frac{\Delta Q}{N_{\mathrm{a}}}, \label{equation_delta_mu}
\end{align}
where $\langle Q \rangle'$ is the order parameter of the interfacial system averaged over the biased MD simulation. $\Delta Q$ is the difference in the order parameter between the solid and liquid. The melting temperature is determined as the point where $\Delta\mu$ becomes zero by using Newton's root finding method. As order parameters, the collective density proposed in Ref.~[\onlinecite{4}] is adopted. Further details of the used parameters and interfacial systems are summarized in Section S4 in the SM~\cite{SM}.

Once the melting temperature has been calculated, one can obtain the entropy of fusion $S_{\mathrm{ls}}$ from the difference in the enthalpy between two phases at the melting temperature. Furthermore, the slope of the melting curve $dT_{\mathrm{m}}/dp$ can be calculated as $S_{\mathrm{ls}}/V_{\mathrm{ls}}$ by the Clausius-Clapeyron relation, where $V_{\mathrm{ls}}$ denotes the volume difference between the liquid and solid. These thermodynamic properties were also evaluated and compared with previously reported values as well as experimental results.

Although the generated force fields are very precise, they necessarily deviate from the FP data as shown in Fig.~\ref{fig6}. The effects of these errors on the melting points were also evaluated by thermodynamic perturbation theory. Details of the thermodynamic perturbation method are described in Ref.~[\onlinecite{9}]. From both the liquid and solid trajectories obtained by 100 ps MD simulations with the MLFF at the calculated melting temperatures, 500-1000 structures are selected.
FP calculations on the selected structures are performed, and the energy difference between the FP and ML potentials are used in thermodynamic perturbation theory. In these calculations, the same supercells and $\mathbf{k}$-points as for the training simulations are employed. For Si, Ge and Sn, we also performed thermodynamic perturbation theory to $3\times 3\times 3$ $\mathbf{k}$-points in order to obtain more accurate results and, for Si, to allow for direct comparison with previous literature values~\cite{34}.

\section{Results: melting points}
\label{section_Tm}
Table~\ref{table_1} summarizes the melting points $T_{\mathrm{m}}$ of Al, Si, Ge, Sn and MgO with and without the thermodynamic perturbation corrections. The entropy of fusion, volume change and slopes of the melting curves calculated by the force fields are summarized in Tables~\ref{table_2}. The uncorrected melting temperatures of Al (LDA and PBEsol), Si (LDA, PBE and SCAN) and MgO (PBE and SCAN) already agree well with the reported DFT results. The thermodynamic perturbation corrections improve the agreement further. We observe that the perturbational corrections (difference between MLFF and CORR in Table~\ref{table_1} ) are more strongly correlated  with the errors in the total energy than with the errors in forces and the stress tensor, indicating that accurate predictions of the total energies are essential for the predictions of the melting points. For Si (LDA and PBE) and MgO (PBE and SCAN), the entropies of fusion, volume changes at the phase transition temperature and slopes of the melting curves also agree well with the recently published reference values\cite{34,35}. These results also indicate that our on-the-fly scheme can efficiently generate force fields applicable to the quantitative predictions of thermodynamic properties. 
\begin{table*}
\caption{Entropies of fusion, $S_{\mathrm{ls}}$ ($k_\mathrm{B}$), volumetric changes, $\Delta V_\mathrm{m} = V_\mathrm{l} - V_{s}$ ($\mathrm{\AA}^{3}\mathrm{atom}^{-1}$), or its relative value to the volume of solid, $V_\mathrm{s}$, and slopes of the melting curves  $dT_\mathrm{m}/dp$ ($\mathrm{K}\mathrm{GPa}^{-1}$) for Al, Si, Ge, Sn and MgO. Errors in our results are within $\pm 0.03$ ($k_{\mathrm{B}}$) for $S_{\mathrm{ls}}$, $\pm 0.02$ ($\mathrm{\AA}^{3}\mathrm{atom}^{-1}$) for $\Delta V_\mathrm{m}$, $\pm 0.003$ for $\Delta V_\mathrm{m}/V_\mathrm{s}$ and $\pm 3$ ($\mathrm{K}\mathrm{GPa}^{-1}$) for $dT_\mathrm{m}/dp$. Recent reference calculations with tight error tolerance  and similar  PAW potential are underlined.}
\label{table_2}
\begin{center}
\begin{tabular}{m{7mm} m{15mm} m{15mm} m{15mm} m{27mm} m{12mm}|m{7mm} m{15mm} m{15mm} m{10mm} m{10mm} m{10mm}}
\hline
        &Property &XC &MLFF &DFT &Exp & &Property &XC &MLFF &DFT &Exp \\
\hline
Si &$S_{\mathrm{ls}}$                      &LDA    &$3.18$   &$3.0^{a}$, $3.5^{b}$, $3.5^{c}$ &$3.3^{d}$    &Al &$S_{\mathrm{ls}}$                      &LDA    &$1.31$   &$1.36^{e}$  &$1.38^{f}$    \\
   &                                       &PBE    &$3.36$   &$\underline{3.3}^{g}$           &$3.6^{h}$    &   &                                       &PBE    &$1.33$   &            &              \\
   &                                       &PBEsol &$3.13$   &                                &             &   &                                       &PBEsol &$1.27$   &            &              \\
   &                                       &SCAN   &$3.47$   &$\underline{3.3}^{g}$           &             &   &                                       &SCAN   &$1.25$   &            &              \\
   &$\Delta V_{\mathrm{ls}}/V_{s}$         &LDA    &$-0.151$ &$-0.10^{a}$, $-0.142^{b}$       &$-0.119^{i}$ &   &$\Delta V_{\mathrm{ls}}$               &LDA    &$1.20$   &$1.26^{e}$  &$1.24^{e}$    \\
   &                                       &PBE    &$-0.128$ &$\underline{-0.120}^{g}$        &$-0.095^{d}$ &   &                                       &PBE    &$1.11$   &            &              \\
   &                                       &PBEsol &$-0.172$ &                                &             &   &                                       &PBEsol &$1.25$   &            &              \\
   &                                       &SCAN   &$-0.089$ &$\underline{-0.091}^{g}$        &             &   &                                       &SCAN   &$1.11$   &            &              \\
   &$dT_{\mathrm{m}}/dp$                   &LDA    &$-69$    &$-50^{a}$, $-58^{b}$, $-51^{c}$ &$-38^{k}$    &   &$dT_{\mathrm{m}}/dp$                   &LDA    &$67$     &            &$65^{j}$      \\
   &                                       &PBE    &$-57$    &$-55^{g}$                       &             &   &                                       &PBE    &$68$     &            &              \\
   &                                       &PBEsol &$-81$    &                                &             &   &                                       &PBEsol &$63$     &            &              \\
   &                                       &SCAN   &$-38$    &$-40^{g}$                       &             &   &                                       &SCAN   &$64$     &            &              \\
\hline
Ge &$S_{\mathrm{ls}}$                      &LDA    &$3.30$   &            &$3.7^{k}$    &MgO     &$S_{\mathrm{ls}}$                      &LDA    &$1.58$   &                                     &              \\
   &                                       &PBE    &$3.50$   &            &             &        &                                       &PBE    &$1.57$   &$\underline{1.62}^{l}$                           &              \\
   &                                       &PBEsol &$3.38$   &            &             &        &                                       &PBEsol &$1.57$   &                                     &              \\
   &                                       &SCAN   &$3.53$   &            &             &        &                                       &SCAN   &$1.50$   &$\underline{1.70}^{l}$                           &              \\
   &$\Delta V_{\mathrm{m}}/V_{\mathrm{s}}$ &LDA    &$-0.124$ &            &$-0.055^{m}$ &        &$\Delta V_{\mathrm{m}}/V_{\mathrm{s}}$ &LDA    &$0.254$  &                                     &              \\
   &                                       &PBE    &$-0.111$ &            &             &        &                                       &PBE    &$0.297$  &$\underline{0.305}^{l}$                          &              \\
   &                                       &PBEsol &$-0.125$ &            &             &        &                                       &PBEsol &$0.267$  &                                     &              \\
   &                                       &SCAN   &$-0.091$ &            &             &        &                                       &SCAN   &$0.269$  &$\underline{0.291}^{l}$                          &              \\
   &$dT_{\mathrm{m}}/dp$                   &LDA    &$-63$    &            &$-20^{k}$    &        &$dT_{\mathrm{m}}/dp$                   &LDA    &$123$    &                                     &              \\
   &                                       &PBE    &$-57$    &            &$-38^{m}$    &        &                                       &PBE    &$153$    &$\underline{153}^{l}$                            &              \\
   &                                       &PBEsol &$-63$    &            &             &        &                                       &PBEsol &$137$    &                                     &              \\
   &                                       &SCAN   &$-44$    &            &             &        &                                       &SCAN   &$140$    &$\underline{134}^{l}$                            &              \\
\hline
Sn &$S_{\mathrm{ls}}$                      &SCAN   &$1.81$   &            &$1.7^{h}$    &        &                                       &       &         &                                     &              \\
   &$\Delta V_{\mathrm{m}}/V_{\mathrm{s}}$ &SCAN   &$0.039$  &            &$0.023^{h}$  &        &                                       &       &         &                                     &              \\
   &$dT_{\mathrm{m}}/dp$                   &SCAN   &$45$     &            &$27^{n}$     &        &                                       &       &         &                                     &              \\
\hline
\multicolumn{12}{l}{\footnotesize{$^a$ Data from Ref.~[\onlinecite{1}].  $\: \: \qquad \qquad$  $^b$ Data from Ref.~[\onlinecite{3}].  $\: \: \qquad \qquad$  $^c$ Data from Ref.~[\onlinecite{4}].  $\: \: \qquad \qquad$  $^d$ Data from Ref.~[\onlinecite{41}].}} \\
\multicolumn{12}{l}{\footnotesize{$^e$ Data from Ref.~[\onlinecite{2}].  $\: \: \qquad \qquad$ $^f$ Data from Ref.~[\onlinecite{46}].  $\: \qquad \qquad$ $^g$ Data from Ref.~[\onlinecite{34}]. $\qquad \qquad$ $^h$ Data from Ref.~[\onlinecite{38}].}} \\
\multicolumn{12}{l}{\footnotesize{$^i$ Data from Ref.~[\onlinecite{42}]. $\qquad \qquad$ $^j$ Data from Ref.~[\onlinecite{47}]. $\qquad \qquad$ $^k$ Data from Ref.~[\onlinecite{36}]. $\qquad \qquad$ $^l$ Data from Ref.~[\onlinecite{35}].}} \\
\multicolumn{12}{l}{\footnotesize{$^m$ Data from Ref.~[\onlinecite{43}].}} \\
\multicolumn{12}{l}{\footnotesize{$^n$ The melting curve of tin was calculated from the volume of $\beta$-tin at 453 K reported in Ref.~[\onlinecite{44}], the volume change by the fusion}} \\
\multicolumn{12}{l}{\footnotesize{$\: \: \:$ written in Ref.~[\onlinecite{45}], and the heat of fusion and melting temperature reported in Ref.~[\onlinecite{38}].}}
\end{tabular}
\end{center}
\end{table*}
\begin{figure}
\includegraphics[width=0.40\textwidth,angle=0]{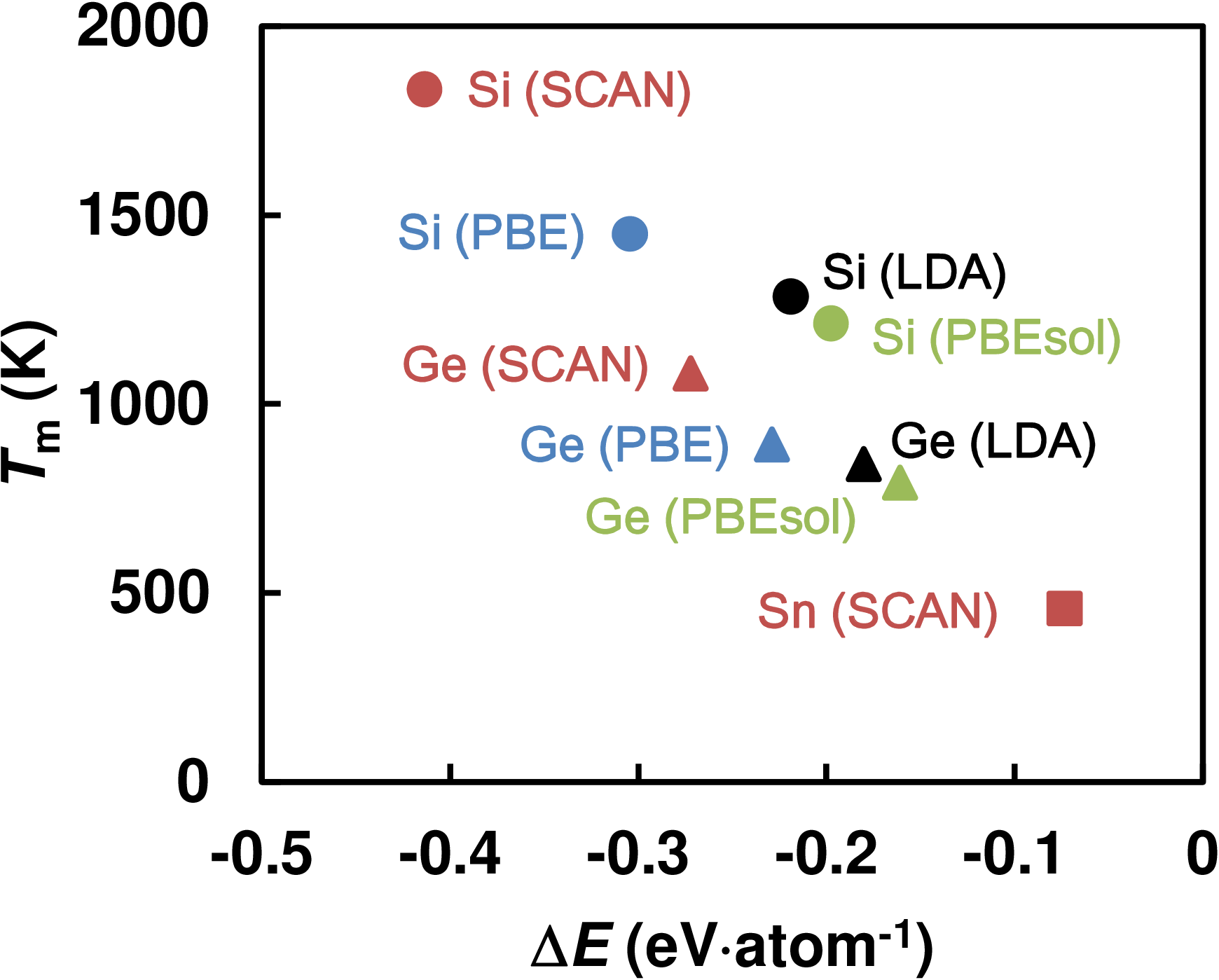}
\caption{Melting point $T_{\mathrm{m}}$ calculated by the interface pinning method using the machine learning force fields and the thermodynamic perturbation corrections vs. the energy difference $\Delta E$ between the $\alpha$- and $\beta$-tin structures obtained from DFT calculations at 0 K. $\Delta E$ is defined to be negative when the $\alpha$-tin structure is more stable than the $\beta$-tin structure.}
\label{fig7}
\end{figure}

Comparison of the theoretical melting points with the experimental results indicates that LDA significantly underestimates the melting point of Si. PBE improves the calculated value, but the melting point is still too low compared to experiment. SCAN, while slightly overestimating the melting point, provides the best agreement with experiment. The worst agreement compared with experiment is obtained by PBEsol, which strongly underestimates the melting point. The trend among PBE, PBEsol and SCAN for Ge is similar to that observed for Si, with the only exception that now even SCAN underestimates the melting point. 

As opposed to Si and Ge, Sn crystallizes in the $\beta$-tin structure. For Sn, PBE and PBEsol do not provide a stable solid within a reasonable temperature range compared to experiment; the melting point seems to be placed at way too low temperatures. Hence, the melting point of Sn has only be determined  for SCAN, which still underestimates the melting temperature. As shown in Fig.~\ref{fig7}, the observed trend among different functionals and materials can be reasonably well correlated to the energy difference between the $\alpha$-tin (the cubic diamond structure) and the $\beta$-tin structures. A similar correlation was already observed in the melting point studies of Si in previous publications~\cite{3,34}, where the trends in the melting points among different functionals were well described by the trends in the energy differences. This is because the energy and structure of liquid Si and Ge can be qualitatively described by the six-fold coordinated structure observed in the $\beta$-tin structure. The results summarized in Fig.~\ref{fig7} indicate that this empirical rule is also roughly applicable to Ge. For Ge, the energy difference between the $\alpha$-tin and $\beta$-tin structure is significantly smaller, and thus, its melting temperature is lower than that of Si. In the case of Sn, the $\alpha$-tin structure is less stable than the $\beta$-tin structure, and melting
is obviously from the $\beta$-tin structure  itself. This implies that the energy difference between both structures might not have a direct relevance for the melting temperature,
nevertheless, the linear relation between the melting temperature and energy difference still seems to apply approximately.

For Al, LDA, PBEsol and SCAN closely reproduce the experimental melting point while PBE underestimates it. Similarly for MgO, LDA, PBEsol and SCAN reproduce the experimental melting point well while PBE underestimates it.

In summary, SCAN is judged to provide the most balanced accuracy within the tested functionals for thermodynamic properties of metallic, covalent and ionic materials. Semilocal functionals always yield lower melting temperatures, but whether LDA, PBEsol or PBE performs better is not {\em a priori} clear. For the tetrahedrally coordinated semiconductors LDA and PBEsol are worse (obviously they underbind the diamond structure), whereas for densely packed materials such as Al, the melting temperature is increased towards the experiment. Since LDA and PBEsol increase the binding energy on average, an increase of the melting temperature for phase transitions where the local structure changes little  is in line with what one would expect. 

\section{Conclusion}
\label{section_conclusion}

An on-the-fly MLFF generation method has been developed and integrated into an electronic structure code. In the developed method, the machine predicts not only the energy, forces and the  stress tensor but also the uncertainty on the basis of Bayesian inference. The predicted error is used to decide whether FP calculations are required or can be bypassed. This error estimation and decision scheme enhances the self-learning ability in the MLFF generation and dramatically reduces the need for human intervention and supervision. The developed method was applied to the calculation of melting points of Al, Si, Ge, Sn and MgO. The application demonstrates that the on-the-fly method indeed enables the efficient generation of accurate MLFFs for metallic, covalent and ionic materials. The MD simulations are more than two orders of magnitude accelerated by the on-the-fly scheme even during learning. Furthermore, for large unit cells, the generated force fields are more than three orders of magnitude faster for MD simulations than FP calculations. This made it possible to calculate the melting points of five materials using the interface pinning method in a fraction of the compute time that would have been required without ML. The melting temperatures predicted by the MLFFs already agree well with the FP results, but it is straightforward and involves only little overhead to employ thermodynamic perturbation theory and correct for the remaining errors. Our on-the-fly method is universally applicable to a wide variety of multi-element complex materials. We believe that this has the potential to become a new working paradigm in the materials science community.

Concerning melting temperatures, we observe that the SCAN functional is clearly a step forward in accurate predictions compared to experiments.
SCAN consistently outperforms the semi-local functionals tested in the present work.  A general trend between the semi-local functionals
can not be made out. For Al, where the local structure remains 12-fold coordinated upon melting, the melting temperature increases from
PBE over LDA to PBEsol towards experiment, in line with the increased cohesive energies predicted by PBEsol and LDA. For Si and Ge, however, the melting temperature 
decreases from PBE over LDA to PBEsol away from experiment. This trend is related to a destabilization of the cubic diamond structure by PBEsol. Clearly, melting temperatures
are a tough test for the performance of density functionals, and only SCAN is quite satisfactory in this regard.

\begin{acknowledgements}
All authors gratefully thank Ryoji Asahi for his many  suggestions on the application and use of machine-learning methods to materials sciences.
\end{acknowledgements}
\appendix
\section{Radial basis functions}
The number of the radial basis functions $N^{l}_{\mathrm{R}}$ is automatically determined such that 
a linear combination of the radial basis functions reproduces 
the radial functions $f_{lm}\left(r,r_{ij}\right)$ with a predefined accuracy. The functions $f_{lm}\left(r,r_{ij}\right)$ are obtained by 
expanding the broadened atomic distribution $\rho_{i}$ [see Eqs.~(\ref{equation_single_atom_distribution}) and (\ref{equation_probability_density})]
in products of spherical harmonics and the  radial functions as
\begin{align}
\rho_{i}\left(\mathbf{r}\right) &= \sum\limits_{l=0}^{L_{\mathrm{max}}} \sum\limits_{m=-l}^{l} \sum\limits_{j=1}^{N_{\mathrm{a}}} f_{lm}\left(r,r_{ij}\right)  Y^*_{lm}\left(\hat{\mathbf{r}}_{ij}\right) Y_{lm}\left(\hat{\mathbf{r}}\right), \label{equation_angular_expansion2} \\
f_{lm}\left(r,r_{ij}\right) &= \frac{4\pi}{\left(\sqrt{2\sigma_\mathrm{atom}^{2}\pi}\right)^{3}} f_\mathrm{cut}\left(r_{ij}\right)  \nonumber \\
&\times\mathrm{exp}\left(-\frac{r^{2}+r_{ij}^{2}}{2\sigma_\mathrm{atom}^{2}}\right) \iota_{l}\left(\frac{rr_{ij}}{\sigma_{\mathrm{atom}}^{2}}\right). \label{equation_radial_function1}
\end{align}
To derive this equation, the following theorem has been used:
\begin{align}
\mathrm{exp} \left(\frac{-|\mathbf{r}-\mathbf{r}_{ij}|^{2}}{2\sigma_\mathrm{atom}^{2}}\right) &= 4\pi \mathrm{exp} \left(-\frac{r^{2}+r_{ij}^{2}}{2\sigma_\mathrm{atom}^{2}}\right) \nonumber \\
&\times \sum\limits_{l=0}^{L_{\mathrm{max}}} \sum\limits_{m=-l}^{l} \iota_{l}\left(\frac{rr_{ij}}{\sigma_{\mathrm{atom}}^{2}}\right) \nonumber \\
&\times Y^{*}_{lm}\left(\hat{\mathbf{r}}_{ij}\right) Y_{lm}\left(\hat{\mathbf{r}}\right). \label{equation_theorem}
\end{align}
The radial part is expanded in a set of radial basis functions $\chi_{nl}\left(r\right)=j_{l}\left(q_{nl}r\right)$ as:
\begin{align}
 \sum\limits_{j=1}^{N_{\mathrm{a}}} f_{lm}\left(r, r_{ij}\right) &= \sum\limits_{n=1}^{N^{l}_{\mathrm{R}}} c_{nlm}^{i} \chi_{nl}\left(r\right). \label{equation_radial_expansion}
\end{align}
Here the parameters $q_{nl}$ are set such that $j_{l}\left(q_{nl}R_{\mathrm{cut}}\right)=0$. 
The number of the radial basis functions $N^{l}_{\mathrm{R}}$ is determined 
to satisfy Eq.~(\ref{equation_radial_expansion}) within a desired precision. 
In our implementation, in advance of the MD simulation, the radial function $f_{lm}\left(r, r_{ij}\right)$ 
is calculated using Eq.~(\ref{equation_radial_function1}) on 100 radial grid points for $r$ 
ranging from 0 to $R_{\mathrm{cut}}$ and $r_{ij}$ ranging from 0.5 $\mathrm{\AA}$ to $R_{\mathrm{cut}}$. 
The number of radial basis functions $N^{l}_{\mathrm{R}}$ is determined to reproduce 
the original values of $f_{lm}\left(r,r_{ij}\right)$ within an error of $\pm0.02$. 
This means that the width of the broadening $\sigma_\mathrm{atom}$ in Eq. (\ref{equation_probability_density}) alone determines
the number of basis functions.

\section{Assumptions necessary for deriving the posterior distributions}
The formulation of the posterior distributions $p\left( \mathbf{w} | \mathbf{Y} \right)$ and $p \left( \mathbf{y} | \mathbf{Y} \right)$ starts from two assumptions:
\begin{enumerate}[i)]
\item{ The FP data vector $\mathbf{y}^{\alpha}$ deviates from the model vector $\bm{\phi}^{\alpha} \mathbf{w}$.
The distribution of the deviation is assumed to be described by a Gaussian function with a covariance matrix
of $\sigma_{\mathrm{v}}^{2}\mathbf{I}$:
\begin{align}
p\left( \mathbf{Y} | \mathbf{w} \right) = \mathcal{N} \left( \bm{\Phi} \mathbf{w}, \sigma_{\mathrm{v}}^{2}\mathbf{I} \right).
\end{align}
$p\left( \mathbf{Y} | \mathbf{w} \right)$ denotes the probability to observe the FP data $\mathbf{Y}$
after the determination of the coefficients $\mathbf{w}$.}
\item {The prior probability to find the vector $\mathbf{w}$ is assumed to be described by a Gaussian distribution with a mean vector at zero and a covariance matrix of $\sigma_{\mathrm{w}}^{2}\mathbf{I}$:}
\begin{align}
p\left( \mathbf{w} \right) &= \mathcal{N} \left( \mathbf{0},\sigma_{\mathrm{w}}^{2}\mathbf{I} \right).
\end{align}
\end{enumerate}
On the basis of these two assumptions and the Bayesian theorem, the posterior distribution $p\left( \mathbf{w} | \mathbf{Y} \right)$ is derived as \
\begin{align}
p\left( \mathbf{w} | \mathbf{Y} \right) &= \frac{p\left( \mathbf{Y} | \mathbf{w} \right) p\left( \mathbf{w} \right)}{p\left( \mathbf{Y} \right)}, \label{equation_Bayesian_theorem} \\
p\left( \mathbf{Y} \right) =& \int p\left( \mathbf{Y} | \mathbf{w} \right) p\left( \mathbf{w} \right) d\mathbf{w}.
\end{align}
The equation can be converted to the Gaussian distribution written as Eq.~(\ref{equation_posterior_distribution_w}) by the completing square method~\cite{26}.
From this posterior distribution, another posterior distribution $p \left( \mathbf{y} | \mathbf{Y} \right)$ is obtained as
\begin{align}
p \left( \mathbf{y} | \mathbf{Y} \right) = \int p \left( \mathbf{y} | \mathbf{w} \right) p \left( \mathbf{w} | \mathbf{Y} \right) d\mathbf{w}.
\end{align}
Similarly to $p\left( \mathbf{w} | \mathbf{Y} \right)$, this distribution can be converted to a Gaussian distribution specified in Eq.~(\ref{equation_posterior_distribution_y})~\cite{26}.

\section{Maximization of evidence function}
The maximization of the evidence function Eq.~(\ref{equation_likelihood}) with respect to the parameters $\sigma_{\mathrm{v}}^{2}$ and $\sigma_{\mathrm{w}}^{2}$ is carried out by simultaneously solving the following equations derived from $\partial p/\partial \sigma_{\mathrm{v}}^{2}=\partial p/\partial \sigma_{\mathrm{w}}^{2}=0$~(see Ref.[\onlinecite{26}]):
\begin{align}
\sigma^{2}_{\mathrm{w}}&=\frac{|\mathbf{\bar{w}}|^{2}}{\gamma}, \label{equation_ea1} \\
\sigma^{2}_{\mathrm{v}}&=\frac{|\mathbf{T}-\bm{\phi}\mathbf{\bar{w}}|^{2}}{M-\gamma}, \label{equation_ea2} \\
\gamma&=\sum\limits_{k=1}^{N_{\mathrm{B}}} \frac{\lambda_{k}}{\lambda_{k}+1/\sigma^{2}_{\mathrm{w}}}. \label{equation_ea3}
\end{align}
Here $\lambda_{k}$ are the eigenvalues of the matrix $\bm{\Phi}^{\mathrm{T}}\bm{\Phi}/\sigma^{2}_{\mathrm{v}}$. As described in Ref.~[\onlinecite{29}], if all eigenvalues are used in the actual computations of Eqs.~(\ref{equation_ea1}), (\ref{equation_ea2}) and (\ref{equation_ea3}), numerical instabilities happen because the non-regularized matrix $\bm{\Phi}^{\mathrm{T}}\bm{\Phi}/\sigma^{2}_{\mathrm{v}}$ can become non-positive definite. In order to avoid this problem, eigenvalues smaller than $10^{-10}$ are excluded from the calculations.

\section{CUR algorithm}
In the following, we denote the element $K(i,j)$ in Eq.~(\ref{equation_U_smeard_expansion}) as $K_{ij}$ 
and the matrix comprised of $K_{ij}$ for all candidates of the local reference configurations as $\mathbf{K}$ 
(both the column and row dimensions of $\mathbf{K}$ are equal to the number of the candidates). 
The formulation of the CUR algorithm starts from the diagonalization of the matrix $\mathbf{K}$. 
\begin{align}
\mathbf{U}^{\mathrm{T}}\mathbf{K}\mathbf{U}&=\mathbf{L}=\mathrm{diag}\left(l_{\mathrm{1}},...,l_{N_{\mathrm{B}}}\right), \label{equation_diag}
\end{align}
where $\mathbf{U}$ is the eigenvector matrix defined as
\begin{align}
\mathbf{U}&=\left(\mathbf{u}_{\mathrm{1}},...,\mathbf{u}_{N_{\mathrm{B}}}\right), \label{equation_U}\\
\mathbf{u}^{\mathrm{T}}_{j}&=\left(u_{1j},...,u_{N_{\mathrm{B}}j}\right), \label{equation_u}
\end{align}
By using the notations in Eq.~(\ref{equation_U}), Eq.~(\ref{equation_diag}) can be rewritten as follows
\begin{align}
\mathbf{k}_{j}&=\sum\limits_{\xi = 1}^{N_{\mathrm{B}}} \left(u_{j \xi} l_{\xi}\right) \mathbf{u}_{\xi}, \label{equation_kj}
\end{align}
where $\mathbf{k}_{j}$ denotes the $j$th column vector of the matrix $\mathbf{K}$. In the original CUR algorithm~\cite{31}, 
the columns of the matrix $\mathbf{K}$ are maintained when they are strongly correlated to the eigenvectors $\mathbf{u}_{\xi}$ with large eigenvalues $l_{\xi}$.
This algorithm was originally developed to efficiently select a few significant local reference configurations 
from many configurations. However, in our on-the-fly force field generation, we need an efficient algorithm 
that can select few insignificant configurations, because the number of configurations discarded by the 
sparsification is usually small. To this end, we have modified the algorithm. In our implementation, 
we dispose of those columns of $\mathbf{K}$ that are strongly correlated with the $N_{\mathrm{low}}$ eigenvectors $\mathbf{u}_{\xi}$ with the small eigenvalues $l_\xi$. The local configurations corresponding to those columns are disregarded. 
Similarly to the original CUR algorithm, the correlation is measured by the statistical leverage 
scoring determined for each column of $\mathbf{K}$ as 
\begin{align}
\omega_{j}&=\frac{1}{N_{\mathrm{low}}} \sum\limits_{\xi=1}^{N_{\mathrm{B}}} \gamma_{\xi j}, \label{equation_ls1}\\
\gamma_{\xi j}&= \begin{cases} u_{j\xi}^{2}, & \mbox{if } l_{\xi}<\mathrm{10}^{\mathrm{-10}} \\ 0, & \mbox{otherwise} \end{cases}. \label{equation_ls2}
\end{align}

\end{document}